%% file: maffucci_tmc1_twocolumn.tex
\documentclass[twocolumn]{aastex61}
\usepackage{comment,mathtools}
\usepackage[utf8]{inputenc}
\usepackage[T1]{fontenc}
\usepackage{textcomp}
\usepackage{gensymb}
\usepackage{wasysym}
\shorttitle{Astrochemical Methods}
\shortauthors{Maffucci et al.}

\begin{document}

\title{Astrochemical Kinetic Grid Models of Groups of Observed Molecular Abundances: Taurus Molecular Cloud 1 (TMC-1)}

\correspondingauthor{Dominique M. Maffucci}
\email{dmm2br@virginia.edu}

\author[0000-0002-4483-1733]{Dominique M. Maffucci}
\affiliation{Department of Chemistry, University of Virginia,
  P.O. Box 400319,
  Charlottesville, VA 22904}
\author[0000-0003-0640-7787]{Trey V. Wenger}
\affiliation{Department of Astronomy, University of Virginia,
  P.O. Box 400325,
  Charlottesville, VA 22904}
\affiliation{ National Radio Astronomy Observatory,
  520 Edgemont Road,
  Charlottesville, VA 22903}
\author[0000-0003-1837-3772]{Romane Le Gal}
\affiliation{Harvard-Smithsonian Center for Astrophysics,
    60 Garden Street,
    Cambridge, MA 02138}
  \author[0000-0002-4649-2536]{Eric Herbst}
  \affiliation{Department of Chemistry, University of Virginia,
    P.O. Box 400319,
    Charlottesville, VA 22904}
  \affiliation{Department of Astronomy, University of Virginia,
  P.O. Box 400325,
  Charlottesville, VA 22904}

\begin{abstract}

  The emission line spectra of cyanoacetylene and methanol reveal chemical and physical heterogeneity
  on very small ($< 0.1$ pc) scales toward the peak in cyanopolyyne emission in the Taurus
  Molecular Cloud, TMC-1 (CP). We generate grids of homogeneous chemical models using a three-phase
  rate equation approach to obtain all time-dependent abundances spanning the physical
  conditions determined from molecular tracers of compact and extended regions of emission along this
  line of sight. Each time-dependent abundance is characterized by one of four features: a
  maximum/minimum, a monotonic increase/decrease, oscillatory behavior, or
  inertness. We similarly classify the time-dependent agreement between modeled and observed abundances
  by calculating both the root-mean-square logarithm difference and root-mean-square deviation
  between the modeled and observed abundances at every point in our grid models for three
  groups of molecules: (i) a composite group of all species present in both
  the observations and our chemical network G, (ii) the cyanopolyynes C = \{HC$_3$N, HC$_5$N,
  HC$_7$N, HC$_9$N\}, and (iii) the oxygen-containing organic species methanol and
  acetaldehyde S = \{CH$_3$OH, CH$_3$CHO\}. We discuss how the Bayesian
  uncertainties in the observed abundances constrain solutions within the grids of chemical models.
  The calculated best fit times at each grid point for each group are tabulated to reveal the minimum
  solution space of the grid models and the effects the Bayesian uncertainties have on the
  grid model solutions. The results of this approach separate the effect different physical conditions
  and model-free parameters have on reproducing accurately the abundances of
  different groups of observed molecular species.
  
\end{abstract}

\keywords{astrochemistry --- ISM: abundances --- methods: miscellaneous}

\section{Introduction}

Emission line surveys of the dark molecular gas in the Taurus Molecular Cloud (TMC-1) reveal a large
gas-phase molecular constituency containing ionic and neutral species, carbon chain molecules, and
oxygen-containing organic species \citep{pratap1997,ohishi1998,markwick2000,markwick2005,soma2015,gratier2016}.
Along the ridge of
molecular gas extending across 5$'$ $\times$ 15$'$ of sky, emission maps obtained with the
QUARRY focal plane array and FCRAO 14-m antennae ($\theta^{14}_{HPBW} = 59'' - 45''$ at $86 - 116$ GHz)
of several species (e.g. SO$_2$, NH$_3$, and HC$_3$N) show that the peaks in emission vary in
location for each molecule suggesting the existence of chemical and physical heterogeneity on a
scale of 0.04 - 0.03 pc at a distance $d = 140$ pc \citep{pratap1997,markwick2000}.
Toward the peak in cyanopolyyne emission, TMC-1 (CP), emission line maps using data collected by
the Nobeyama 45-m ($\theta^{45}_{HPBW} = 20''$ at $96$ GHz) show 
the spatial separation of the peaks in molecular emission intensities for methanol and carbon monosulfide,
a tracer of the dense molecular clouds, to exist
on a smaller scale of 0.01 pc across 150$''$ $\times$ 150$''$ of sky \citep{soma2015}. The broadband spectral
line survey toward TMC-1 (CP) conducted using the Nobeyama 45-m telescope from 8.8 to 50 GHz
corresponding to $\theta_{HPBW} = 156'' - 27.5''$ and projected linear scale $s = 0.0187 -0.106$ pc at a distance $d = 140$ pc
provides a rich data set enabling the
simultaneous analysis of the emission line spectra of several molecular components \citep{kaifu2004}.
Relative molecular abundances and a corresponding set of uncertainty values has been determined using
these data and a Bayesian analysis of an LTE model of radiative transfer that both detects outlier emission lines with
respect to assumed prior uncertainty distributions and determines the uncertainties in the calculated column densities \citep{gratier2016}.

The line of sight toward TMC-1 (CP) has been well studied in several emission line surveys, and
homogeneous chemical models with cold ($T = 10$ K), dark ($n \sim 10^5$ cm$^{-3}$, $A_{{\rm V}} \sim 10$) molecular cloud conditions reproduce large sets
(> 50) of observed relative molecular abundances within an average factor of ten
\citep{wakelam2006,garrod2007,agundez2013,loison2014,ruaud2015,ruaud2016}. Homogeneous chemical
models worsen, however, for increasingly large sets of observed abundances because
within a group of molecular species, the dominant production and destruction reaction mechanism
sequences are unique to each molecule and vary in time.
For example, species with large energies of desorption
or lacking efficient gas-phase formation mechanisms (e.g. methanol, CH$_3$OH) require
additional mechanisms to gas-phase reactions, such as grain-surface processes, to reproduce the
observed gas phase abundances in homogeneous models of dark cloud conditions
\citep{garrod2007,vasyunin2013,ruaud2015}. With additional gas-grain processes like the Eley-Rideal
and van der Waals complex-induced reaction mechanisms \citep{ruaud2015}, nonthermal desoption mechanisms
like that due to the exothermicity of surface reactions \citep{garrod2007,minissale2016} enrich
the cold, dense gas with large amounts of oxygen-containing organic species methanol
and acetaldehyde, so chemical models that incorporate these new mechanisms
should benchmark the effects of their addition to the chemical network against both observed molecular
abundances and existing chemical models of cold, dark cloud conditions.

Because different chemical components appear to have dissimilar physical conditions along
the line of sight to TMC-1 (CP) and because molecules are
sensitive both to grain and ice surface processes and non-thermal
desorption, we compute grids of homogeneous chemical models over ranges of free parameters
such as the physical conditions density $n$ and cosmic-ray ionization rate $\zeta$
as well as the reactive-desorption efficiency $a$ and the diffusion-binding
energy ratio $b$, which control the grain-surface reaction mechanisms.
Using two elemental compositions for models with and without activated Eley Rideal and van der Waals
complex reaction mechanisms, we create four distinct grid models to reveal the effects these additional parameters
have on the model of the time-dependent chemical
structure of TMC-1 (CP). Next, we quantify the agreement between each model and sets of observed
abundances toward TMC-1 (CP) by calculating the root-mean-square logarithm difference between
the observed and modeled abundances \citep{wakelam2006,hincelin2011,vidal2017} at every time and for every combination of free
parameters for three groups of molecular abundances: a
composite Group G = \{NH$_3$, CH$_3$OH, c-C$_3$H, l-C$_3$H, l-C$_3$H$_2$, c-C$_3$H$_2$, CH$_3$CCH,
H$_2$CCN, CH$_3$CN, H$_2$CCO, CH$_3$CHO, HCS$^+$, H$_2$CS, C$_4$H$_2$, C$_3$N, HCCNC, C$_3$O,
HC$_3$NH$^+$, CH$_2$CHCN, C$_5$H, CH$_3$C$_4$H, CH$_3$C$_3$N, C$_3$S, C$_6$H, HC$_3$N, HC$_5$N,
HC$_7$N, HC$_9$N\} of all molecules contained in both the latest benchmark chemical composition of
TMC-1 \citep{gratier2016} and in our chemical network, Group C = \{HC$_3$N, HC$_5$N, HC$_7$N,
HC$_9$N\} of the four smallest cyanopolyynes, and Group S = \{CH$_3$OH, CH$_3$CHO\} of methanol and acetaldehyde; all
molecular abundance differences contribute equally to the group mean and are weighted equally using this method.
Additionally for each group, we calculate the root-mean-squared deviation of the
modeled abundances with respect to the observed abundances using the Bayesian uncertainties \citep{gratier2016}
so that each difference in logarithm abundance is weighted by its observed uncertainty, and we
compare these weighted fits with the unweighted fits. Finally, we identify the best models for each group and
discuss both the effect the molecular group has on the grid model solution region and how
uncertainties in the observed abundances constrain the model fits accordingly.

\section{Chemical Model}

\subsection{Physical Conditions}

The dipole-allowed transition selection rules $\Delta J = \pm 1$ and $\Delta K = 0$ (where $J$
and $K$ are quantum numbers for the total angular momentum and its projection on the
axis of molecular symmetry) of symmetric top molecules like methyl acetylene CH$_3$CCH
separate the effects of radiative and collisional excitation. The
kinetic temperature of the gas is reflected in the relative total
populations of all $J$ levels in each $K$ ladder, which is controlled exclusively by
collisional excitation \citep{bergin1994,pratap1997}. Toward TMC-1 (CP), the temperature
measured using methyl acetylene emission ($J = 6 \rightarrow 5, K = 0, 1, 2$)
and a statistical equilibrium analysis in the LVG approximation is found to be 10 K and is consistent with the temperature determined
from other molecular tracers of kinetic temperature like ammonia \citep{pratap1997}. We adopt
this temperature, which has been used in previous TMC-1 (CP) models, as the only value for
both the gas and grain temperatures in our chemical model grids.
Futhermore, statistical equilibrium analysis of the excitation of cyanoacetylene HC$_3$N
($J = 4 \rightarrow 3, J = 10 \rightarrow 9,$ and $J = 12 \rightarrow 11$) constrains the
density to $n = 8 \times 10^4$ cm$^{-3}$ at temperature 10 K toward TMC-1 (CP) \citep{pratap1997}.

The total nuclear spin $I$ of the three hydrogen nuclei contained in the internally rotating methyl group (-CH$_3$)
distinguishes either symmetric A$^{\pm}$ ($I = 3/2,\ ortho$) or antisymmetric E ($I = 1/2,\ para$)
nuclear spin states, each of which combines with rotational states that are antisymmetric or
symmetric, respectively, to form a distinct rotational energy level spectrum \citep{sutton1988,rabli2010a,rabli2010b,levshakov2011}.
Following the
$a$-type transition selection rules for asymmetric tops $\Delta J = -1, \Delta K_A = 0,$ and
$\Delta K_C = -1$, maps of the spatially resolved transitions
$J_{K_A,K_C} = 1_{0,1} \rightarrow 0_{0,0}, 2_{0,2} \rightarrow 1_{0,1}, 3_{0,2} \rightarrow 2_{0,1}$
for A$^+$ states and $J_{K_A,K_C} = 2_{-1,2} \rightarrow 1_{-1,1},$ and $3_{-1,2} \rightarrow 2_{-1,1}$
for E states reveal lower densities $n = 1 - 4\ \times\ 10^4$ cm$^{-3}$ compared with the
cyanopolyyne emission along the line of sight to TMC-1 (CP) suggesting that
methanol emission arises from an extended region of lower density \citep{soma2015}.
Our grid models vary over three densities, $n = 10^4, 10^{4.5},$ and $10^5$ cm$^{-3}$ (see Table
\ref{tab_grid}), spanning the values fit to the cyanoacetylene \citep{pratap1997} and methanol
\citep{soma2015} emission toward TMC-1 (CP).

Though the observed molecular emission spectra
constrain the kinetic temperatures and densities of each emitting molecular component along the line of sight,
other physical
conditions like the visual extinction, A$_{{\rm V}}$, and cosmic ray ionization rate,
$\zeta$, depend on the total hydrogen column density,
$N_{{\rm H}} = N({\rm H}) + 2N({\rm H}_2)$, which is usually assumed to be constant with respect
to the total column density of another molecular component throughout the emitting column.
Imposing a constant ratio of the molecular hydrogen column density with respect to that of a
molecular tracer to normalize a set of molecular column densities, however, presupposes each
region of molecular emission arises from similar conditions, and this could introduce a
systematic error into a set of abundance values scaled by this method \citep{liszt2000}.
For abundances normalized by $N({\rm H}_2) = 10^{22}$ cm$^{-2}$ \citep{gratier2016}, we select
typical dark cloud values for the central extinction $A_{{\rm V}} = 10$ and
cosmic ray ionization rate $\zeta = 10^{-17}, 10^{-17.5},$ and $10^{-16}$ s$^{-1}$
(see Table \ref{tab_grid}) for consistency with previous studies of the chemical structure of
TMC-1 (CP) \citep{garrod2007,vasyunin2013,ruaud2015,ruaud2016}.

\begin{deluxetable}{lc}
\tabletypesize{\small}
\tablecaption{Model Parameter Space \label{tab_grid}}
\tablehead{ \colhead{Parameter} & \colhead{Value(s)} 
}
\decimals
\startdata
 $T_{{\rm gas}}$ (K) & 10 \\
 $n$ (cm$^{-3}$) & 10$^5$, 10$^{4.5}$, 10$^4$ \\
 $A_{{\rm V}}$ (mag) & 10  \\
 $\zeta$ (s$^{-1}$) & 10$^{-17}$, 10$^{-16.5}$, 10$^{-16}$ \\
 $T_{{\rm dust}}$ (K) & 10  \\
 $a$ & 0.01, 0.03, 0.1 \\
 $b$ & 0.3, 0.4, 0.5 \\
\enddata
\end{deluxetable}

\subsection{Chemical Kinetics}

The ionization of molecular hydrogen by cosmic-ray impact initiates sequences of ion-neutral
reactions, neutral-neutral reactions, and dissociative recombinations with free electrons,
enriching the gas with several generations of molecular species of increasing complexity
\citep{herbst1973,woon1996,smith2004,woon2009}. As gas-phase species accrete onto dust
grain surfaces, light species gain mobility, and grain-surface chemistry proceeds via the
Langmuir-Hinshelwood mechanism as species thermally diffuse \citep{hasegawa1992} or quantum
mechanically tunnel \citep{hasegawa1993a} from binding site to binding site across the grain surface. 
Ice mantles develop when the number of accreted molecules exceeds the number of binding sites
on the surface, and the species contained within these mantles can further react as they
diffuse through the bulk \citep{hasegawa1993b,garrod2011,garrod2013,ruaud2016}. The diffusion
barrier ($E_{{\rm diff}}$) between adjacent binding sites on the ice surface and within the ice mantle is
unique for each molecule and modeled as a fraction of the binding energy of adsorption,
$E_{{\rm diff}} = bE_{{\rm des}}$ where $0 < b < 1$. Several values of the ratio of the surface diffusion
barrier to the binding energy, $b_s = E^s_{{\rm diff}}/E_{{\rm des}}$, have been used in dark cloud models,
and we select three ratios corresponding to a low \citep{hasegawa1992}, moderate, and high
\citep{ruaud2015} value, $b_s = 0.3, 0.4,$ and $0.5$, respectively. For the ratio of the bulk mantle
diffusion barrier to the binding energy, $b_m = E^m_{{\rm diff}}/E_{{\rm des}}$, we use a single value of
0.8 for all species \citep{ruaud2016}.

In previous dark cloud two-phase models
in which only the gas and ice surface were chemically active \citep{garrod2007}, the
large binding energy of methanol inhibited the liberation of grain-surface methanol back
to the gas, and without some non-thermal desorption mechanism, the observed gas-phase abundance
of methanol was underproduced by the chemical models. Both photodesorption
and desorption via exothermic surface reactions are nonthermal mechanisms by which methanol
and other surface-bound species can enrich the gas phase, but the densities and
extinctions in dark cloud models make photodesorption inefficient and exemplify situations
in which reactive desorption specifically controls a gas-phase interstellar molecular abundance \citep{garrod2007,vasyunin2013}.
This study employs the Rice-Ramsperger-Kessel (RRK) formulation of the reactive desorption
probability and parameter $a$ \citep{garrod2007,vasyunin2013}.
A recent semiempirical approach invokes the equipartition of energy and an elastic collision
process to model the probability of reactive desorption as seen in experiments
\citep{minissale2016}. In this formulation, the energy loss due to exothermicity is
transferred to a component perpendicular to the substrate surface, and the efficiency is
scaled by the masses of the product and the surface. This method allows for a greater
level of detail to be considered for each reaction of this type within a chemical network. 

Gas-phase species that collide with grain ice surfaces can also react upon collision via
the Eley-Rideal mechanism or form weakly bound van der Waals complexes, which can then
undergo hydrogenation to saturate; these mechanisms enhance the abundances of
oxygen-containing organic precursors to methanol and acetaldehyde, which then hydrogenate
and sufficiently enrich the gas with nominal reactive desorption efficiency ($a = 0.01$)
\citep{ruaud2015}. Our grid models include nominal, moderate, and high reactive desorption
efficiencies $a = 0.01, 0.03$, and $0.10$ (see Table \ref{tab_grid}).

\subsection{Reaction Network and Rate Solver}

The evolution of the time-dependent molecular volume densities [A] of a group of molecules
A = \{A$_1$, A$_2$, A$_3$, ..., A$_n$\} subject to a system of chemical reactions (the chemical
network) is obtained by integrating the corresponding system of differential rate law
equations for each species $i$:
\begin{eqnarray}\label{eq:rate_law}
  \frac{d\big{[}{\rm A}_i\big{]}}{dt} &=& \sum_j\sum_l k_{jl}\big{[}{\rm B}\big{]}_j\big{[}
    { \rm C}\big{]}_l + \sum_p k_p\big{[{}\rm D}\big{]}_p \nonumber \\
  & &- \big{[}{\rm A}_i\big{]}\Bigg{(}\sum_m k_m\big{[}{\rm F}\big{]}_m + \sum_q k_q\Bigg{)}
\end{eqnarray}
\noindent where the first two terms express the two-body and one-body production mechanisms of A
and the third and fourth terms represent the two-body and one-body
destruction pathways. The abundances are computed by normalizing each time-dependent molecular
volume density by the hydrogen volume density. Our chemical network is a combination of the
latest gas-phase reactions of the periodically updated and benchmarked KInetic Database for
Astrochemistry KIDA \citep{wakelam2012,wakelam2015,ruaud2015,ruaud2016,vidal2017,legal2017} and a
grain-surface-mantle network previously coupled with the KIDA \citep{garrod2007,ruaud2015,ruaud2016}
to form a composite reaction network for use in three-phase (gas, ice-surface, ice-mantle)
chemical models. To integrate
the system of coupled, nonlinear differential rate law equations, we use the
\texttt{Nautilus} code \citep{hersant2009,hincelin2011,hincelin2013,hincelin2016,ruaud2016}, which has recently been updated to include the Eley-Rideal and van
der Waals complex induced reaction mechanisms \citep{ruaud2015} and three-phase capabilities
considering chemistry in the gas, on the ice surface, and throughout the ice mantle
\citep{ruaud2016}.

\subsection{Elemental Composition}

To account for material that is absent from the gas but still contributes
to the total elemental composition of dark molecular clouds, a set of cosmic references abundances,
typically the solar elemental abundances, must be depleted by factors reflecting unobserved ice-phase material as well as the
refractory dust grains, which are composed of amorphous olivine in our models \citep{semenov2010,hincelin2011}.
For the initial fractional elemental abundances with respect to the total
hydrogen density (see Table \ref{tab_elem}), we select the canonical low-metal abundances
\citep{morton1974,graedel1982} and include modifications for helium \citep{wakelam2008},
carbon and nitrogen \citep{jenkins2009}, and fluorine \citep{neufeld2005}. Oxygen, which is contained
in both ice species and the dust grains of dark clouds, has
a depletion factor that has been shown to increase with
increasing density over a sample of hundreds of lines of sight with diffuse cloud densities
$n({\rm H}) \leq 10$ cm$^{-3}$ \citep{jenkins2009}. We use the two values
$f_{{\rm O}} = 2.4 \times 10^{-4}$ and $1.4 \times 10^{-4}$
extrapolated for dense cloud conditions \citep{hincelin2011} representing intermediate and high depletion cases, respectively.
The time-dependent abundances of the cyanopolyynes and small oxygen species are sensitive to the C/O ratio \citep{wakelam2010};
the values
$f_{{\rm C}}/f_{{\rm O}} = 0.7$ and $1.2$
in our models differentiate oxygen-rich, carbon-poor conditions from oxygen-poor,
carbon-rich conditions and illustrate the effect oxygen depletion has on models of dark interstellar clouds and the
abundances toward TMC-1 (CP).

\begin{deluxetable}{lc}
\tabletypesize{\small}
\tablecaption{Fractional Elemental Abundances $f_i = n_i/n_{\rm H}, n_{\rm H} = n({\rm H}) + 2n({\rm H}_2)$ \label{tab_elem}}
\tablehead{\colhead{Element} & \colhead{Abundance} 
}
\decimals
\startdata
 H$_2$ & 0.5 \\
 He & 9.0(-2) \\
 O & 1.4(-4), 2.4(-4)\tablenotemark{*}\\
 N & 6.2(-5)\\
 C$^+$ & 1.7(-4)\\
 S$^+$ & 8.0(-8) \\
                Si$^+$ & 8.0(-9)\\
 Fe$^+$ & 3.0(-9)\\
 Na$^+$ & 2.0(-9) \\
 Mg$^+$ & 7.0(-9) \\
P$^+$ & 2.0(-10) \\
 Cl$^+$ & 1.0(-9) \\
 F & 6.68(-9)\\
\enddata
\tablenotetext{*}{Increasing oxygen abundance corresponds to decreasing oxygen depletion for dark cloud conditions \citep{hincelin2011}.}
\end{deluxetable}

\subsection{Grid Models}

In general, grids of homogeneous chemical kinetic models demonstrate the effect
that variations in the initial parameters of the rate law equations have on the time-dependent
abundance solutions. Methods of mapping homogeneous chemical models to arrays of observationally
constrained physical conditions have reproduced the chemical structure of Active Galactic Nuclei
\citep{harada2013} and protoplanetary disks \citep{oberg2015,cleeves2016}, and chemical
heterogeneity along a single line of sight warrants a similar approach to model the accompanying
physical heterogeneity of the spatially distinct regions of emission. Astrochemical grid models
have also been used to determine time-dependent column densities, benchmark chemical networks
and models, and predict molecular emission line intensities in starburst galaxies from
statistical equilibrium calculations in the LVG approximation \citep{viti2017}. We automate the
execution of the \texttt{Nautilus} code \citep{ruaud2016} over a 7-D parameter space (gas kinetic temperature $T_{{\rm gas}}$, density $n$,
visual extinction $A_{{\rm V}}$, cosmic ray ionization rate $\zeta$, dust temperature
$T_{{\rm dust}}$, reactive desorption efficiency $a$, and diffusion to binding energy ratio $b$),
as shown in Table \ref{tab_grid}, and for each molecule $i$ we construct a 9-D data structure
\{$X_i,t,T_{{\rm gas}},n,A_{{\rm V}},\zeta,T_{{\rm dust}},a,b$\} containing the abundance $X_i$
at every time $t$ and every combination $p$ of free parameters \{$T_{{\rm gas}},n,A_{{\rm V}},\zeta,T_{{\rm dust}},a,b$\}.
To obtain solutions that explicitly separate the effects of the new mechanisms and the
elemental composition, we compute four grids, Models A, B, C, and D, for two elemental
compositions, C/O = 0.7 and 1.2 corresponding to intermediate and high cases of oxygen depletion, and either inactive (N) or active (Y)
Eley-Rideal (ER) and van der Waals (vdW) reaction mechanisms (see Table \ref{mod_grid}).

For each position $(p,t) = (T_{gas},n,A_{{\rm V}},\zeta,T_{dust},a,b,t)$ in our grids and for each
molecular Group $M$ = {G, C, S}, we calculate the root-mean-square (hereafter rms) log difference
\begin{equation}\label{fit_a}
  \mathcal{A}(p,t) = \Bigg{[}\frac{1}{n}\sum_i^n[{\rm log}(X_{{\rm mod}}(p,t)/X_{{\rm obs}})]^2_i\Bigg{]}^{1/2}
\end{equation}
\noindent between the modeled and observed fractional abundances, $X_{{\rm mod}}$ and $X_{{\rm obs}}$,
respectively, for the $n$ molecular abundances in each group. The rms
log difference $\mathcal{A}(t)$ of each group quantifies the average factor of agreement between
the observed and corresponding modeled abundances but neglects uncertainties in both the modeled and
observed abundances.
Meaningful solutions to the rate law equations exist for the observed abundances when
the rms log difference is less than some criterion value $\mathcal{A}_{{\rm crit}}$, 
and we impose $\mathcal{A}_{{\rm crit}} = 1$ corresponding to an average factor of agreement between modeled and observed abundances within
each group of one order of magnitude. The best fit time $\mathcal{T}$ or chemical timescale
of each model with parameters $p$ is calculated by minimizing the rms log
difference $\mathcal{A}_{{\rm min}} = \mathcal{A}(p,\mathcal{T})$ for each model.

Statistical methods, specifically the Bayesian analysis of emission line spectra, however, produce
column densities with uncertainties $\sigma_i$ that reflect the prior uncertainty distributions of free
parameters of the LTE model of the emission spectra of each molecular component \citep{gratier2016}.
As shown in Table \ref{tab_abs}, the uniqueness of each observed molecular emission
spectrum results in unequal standard deviations for the LTE column densities, and these
uncertainties propagate unchanged to the observed abundances if a constant hydrogen column
density is assumed along the regions of integrated emission traced by each of the column densities. For each group
of molecules, the agreement between a modeled set of abundances and an observed set of abundances
with a corresponding set of uncertainties can be quantified by the mean deviation
\begin{equation}\label{fit_sig}
  \sigma(p,t) = \Bigg{[}\frac{1}{n}\sum_i^n\Bigg{(}\frac{{\rm log}(X_{{\rm mod}}(p,t)/X_{{\rm obs}})}{\sigma_i}\Bigg{)}^2_i\Bigg{]}^{1/2}
\end{equation}
\noindent between the modeled and observed abundances in units of $\sigma_i$, the 1$\sigma$ deviations
associated with each of the observed abundances. The rms deviation $\sigma(p,t)$
or weighted fit measure reduces to the unweighted fit measure $\mathcal{A}(p,t)$ when all 1$\sigma$
deviations $\sigma_i$ are unity corresponding to a standard deviation of an order of magnitude
difference between the modeled and observed relative molecular abundances.
Because the 1$\sigma$ values are reported with the observed abundances, we use
$\sigma_{{\rm crit}} = 1$ as the solution criterion similar to the rms log difference.
The weighted fit measure inaccurately expresses the mean deviation for a group of molecules with
individual deviations that largely differ from each
other. As terms diverge in value and some begin to dominate the sum while others diminish in
contribution, the reduction factor $\frac{1}{\sqrt{n}}$ of the size of the group $n$ no longer represents the
number of molecules that meaningfully contribute, and the weighted fit measure $\sigma(p,t)$
underestimates the mean deviation within the group. The same is
true for the unweighted fit measure $\mathcal{A}(p,t)$, which lacks the uncertainties as weights.
Small uncertainties demand better agreement between the modeled and observed abundances for equal
contribution to the mean, and both biases favor the large contributions to the mean.
The observed abundances of molecules in each group C and S exhibit uncertainties $\sigma_i$
that resemble the the rest of the group so that the
mean deviation has meaning when grouping in this manner and corresponds to a similar
factor of ageement between the observed and modeled abundances within each group.

\begin{deluxetable}{lcc}
\tabletypesize{\small}
\tablecaption{Model Elemental Composition and Reaction Mechanism Spaces \label{mod_grid}}
\tablehead{ \colhead{Model} & \colhead{C/O} & \colhead{ER/vdW} 
}
\decimals
\startdata
 A & 0.7 & N \\
 B & 0.7 & Y \\
 C & 1.2 & N \\
 D & 1.2 & Y \\
\enddata
\end{deluxetable}

\section{Results and Discussion}

\begin{deluxetable}{lcccccc}
\tabletypesize{\footnotesize}
\tablecaption{Relative Molecular Abundances $X_i = N_i/N_{{\rm H}}$ and 1$\sigma$ Uncertainties toward TMC-1 CP \citep{gratier2016} for
  Group(s) $M$, and Character Type(s) $\mathcal{C}$ for Restricted Grid Models A, B, C, and D \label{tab_abs}}
\tablehead{ \colhead{Molecule} & \colhead{log($X_i$)} & \colhead{$M$} & \colhead{$\mathcal{C}$(A)}& \colhead{$\mathcal{C}$(B)}&
  \colhead{$\mathcal{C}$(C)}& \colhead{$\mathcal{C}$(D)}}
\decimals
\startdata
NH$_3$ & -7.30$^{+0.61}_{ -2.33}$ & G & 1 & 1 & 1 & 1 \\
CH$_3$OH & -8.84$^{+0.25}_{ -1.79}$ & G, S & 3  & 3 & 3 & 3 \\
c-C$_3$H & -8.52$^{+0.07}_{ -0.05}$ & G & 1 & 1 & 1 & 1 \\
l-C$_3$H & -9.25$^{+0.07}_{ -0.03}$ & G & 1 & 1 & 1 & 1\\
l-C$_3$H$_2$ & -10.23$^{+0.50}_{ -1.90}$ & G& 1 & 1 & 1 & 1 \\
c-C$_3$H$_2$ & -8.73$^{+1.13}_{ -2.90}$ & G & 3 & 3 & 2,3 & 2,3\\
CH$_3$CCH & -7.94$^{+1.16}_{ -1.26}$ & G& 2 & 2 & 2 & 2 \\
H$_2$CCN & -8.42$^{+0.24}_{ -0.34}$ & G & 1& 4 & 2,3 & 2,3\\
CH$_3$CN & -9.39$^{+0.19}_{ -0.18}$ & G& 2 & 2 & 2 & 2 \\
H$_2$CCO & -9.32$^{+0.35}_{-1.71}$ & G & 2 & 2 & 2,3 & 2,3 \\
CH$_3$CHO & -9.57$^{+0.31}_{ -2.03}$ & G, S & 2 & 2 & 2 & 2\\
HCS$^+$ & -9.24$^{+0.53}_{ -0.38}$ & G & 1& 1& 1 & 1\\
H$_2$CS & -8.38$^{+0.53}_{ -0.13}$ & G &3 & 3& 3& 3\\
C$_4$H$_2$ & -8.72$^{+0.23}_{ -0.30}$ & G &3 &2,3 & 2,3& 2,3\\
C$_3$N & -8.45$^{+0.21}_{ -0.30}$ & G & 1,2,3,4 & 1,2,3,4 & 2 & 2 \\
HC$_3$N & -7.63$^{+0.13}_{ -0.06}$ & G, C & 2 & 2 & 2& 2\\
HCCNC & -9.07$^{+0.31}_{ -0.11}$ & G & 2,3,4 & 2,3,4 & 2& 2\\
C$_3$O & -10.08$^{+0.25}_{ -1.29}$ & G &2 & 2& 2& 2\\
HC$_3$NH$^+$ & -10.13$^{+0.45}_{ -1.81}$ & G &2,3,4 & 2,3,4 & 2& 2\\
CH$_2$CHCN & -9.19$^{+0.22}_{ -0.08}$ & G & 2,3&2,3 & 2& 2\\
C$_5$H & -9.73$^{+0.12}_{ -0.10}$ & G & 1,3,4 & 2,3,4 & 2,3&2.3 \\
CH$_3$C$_4$H & -8.83$^{+0.23}_{ -0.17}$ & G & 2 & 4& 2,3& 2,3\\
CH$_3$C$_3$N & -10.01$^{+0.18}_{ -0.17}$ & G &2 &2,3,4 & 2& 2\\
C$_3$S & -8.86$^{+0.16}_{ -0.11}$ & G & 2,3 & 2,4& 2,3& 2,3\\
C$_6$H & -9.26$^{+0.04}_{ -0.05}$ & G & 2 & 2& 2,3& 2,3\\
HC$_5$N & -8.23$^{+0.10}_{ -0.09}$ & G, C & 3,4 & 4& 2,3&2,3 \\
HC$_7$N & -8.34$^{+0.14}_{ -0.10}$ & G, C & 4& 4 & 2,3 & 2,3 \\
HC$_9$N & -8.98$^{+0.06}_{ -0.06}$ & G, C & 4& 4 & 2,3 & 2,3 \\
\enddata
\tablecomments{Character type $\mathcal{C}$ = 1 (inertness), 2 (maximum/minimum), 3 (monotonic increase/decrease), and 4 (oscillatory).}
\end{deluxetable}

\subsection{General Characteristics of the Modeled Abundances}

\begin{figure*}
\plotone{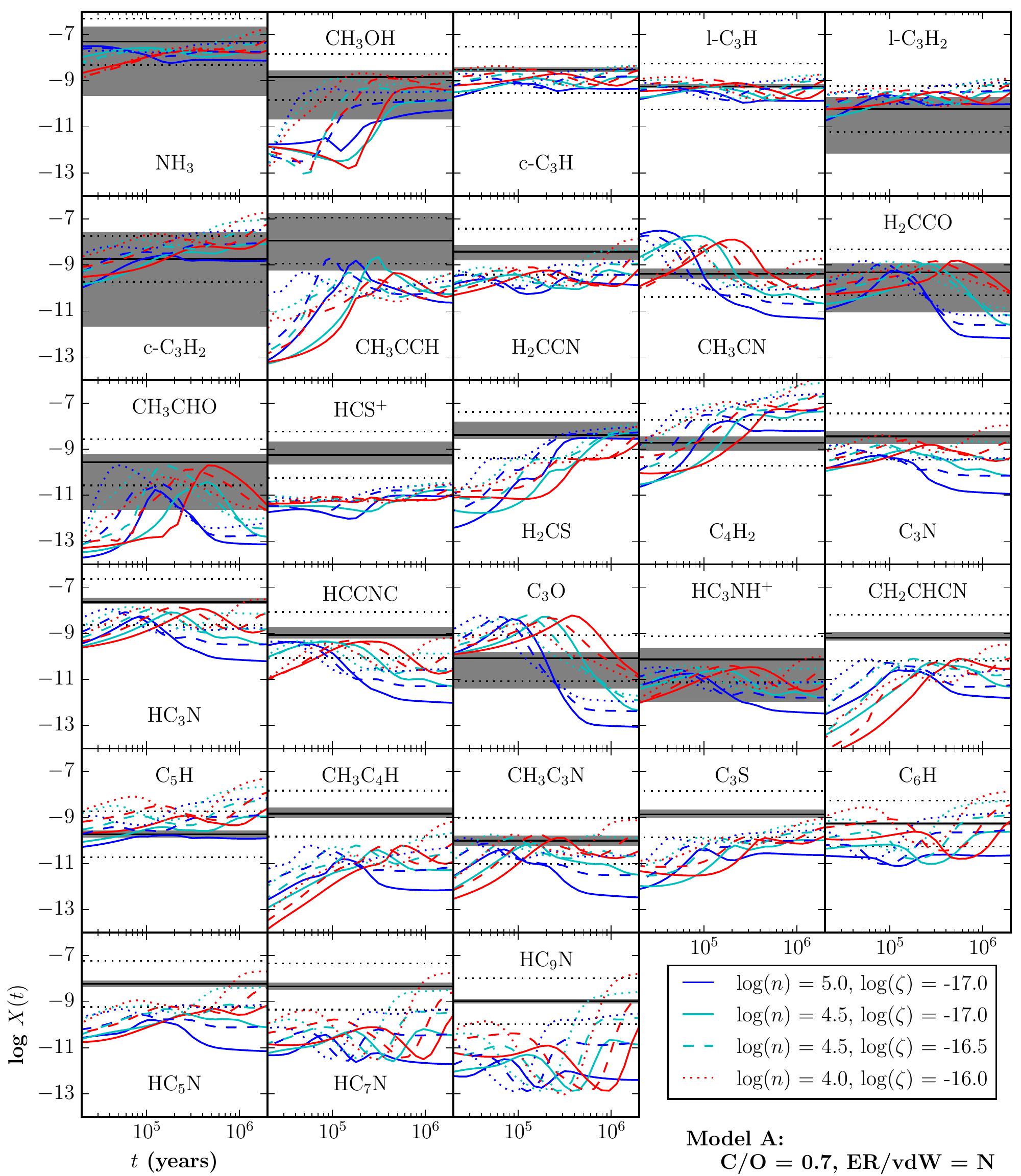}
\caption{Time-dependent abundances log $X(t)$ for Model A (C/O = 0.7, ER/vdW = N, $a$ = 0.01, $b$ = 0.4)
  with observed abundances log $X_{{\rm obs}}$ in solid black, an order of magnitude difference from the observed
  abundances log $X_{{\rm obs}} \pm 1$ in dotted black, and the 1$\sigma$ uncertainties on the observed abundances
  in gray. The linestyle corresponds to the
cosmic ray ionization rate: solid ($\zeta = 10^{-17}\ $s$^{-1}$), dashed ($\zeta = 10^{-16.5}\ $s$^{-1}$), and dot-dashed
($\zeta = 10^{-16}\ $s$^{-1}$), while the color indicates the value of density: blue ($n = 10^5\ {\rm cm}^{-3}$),
cyan ($n = 10^{4.5}\ {\rm cm}^{-3}$), and red ($n = 10^4\ {\rm cm}^{-3}$).} \label{abs_low}
\end{figure*}

\begin{figure*}
\plotone{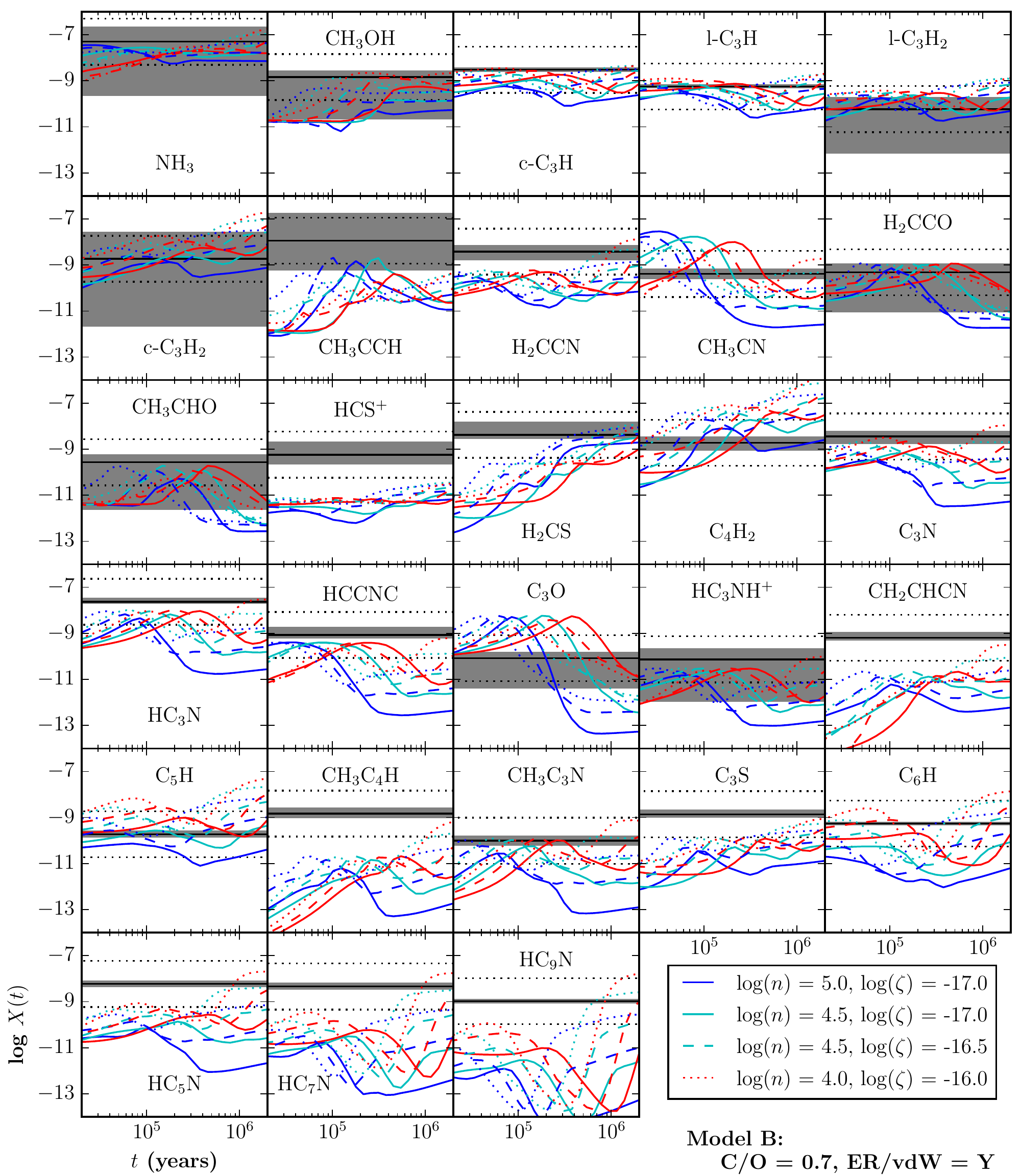}
\caption{Time-dependent abundances log $X(t)$ for Model B (C/O = 0.7, ER/vdW = Y, $a$ = 0.01, $b$ = 0.4)
  with observed abundances log $X_{{\rm obs}}$ in solid black, an order of magnitude difference from the observed
  abundances log $X_{{\rm obs}} \pm 1$ in dotted black, and the 1$\sigma$ uncertainties on the observed abundances
  in gray. The linestyle corresponds to the
  cosmic ray ionization rate: solid ($\zeta = 10^{-17}\ $s$^{-1}$), dashed ($\zeta = 10^{-16.5}\ $s$^{-1}$), and dot-dashed
  ($\zeta = 10^{-16}\ $s$^{-1}$), while the color indicates the value of density: blue ($n = 10^5\ {\rm cm}^{-3}$),
  cyan ($n = 10^{4.5}\ {\rm cm}^{-3}$), and red ($n = 10^4\ {\rm cm}^{-3}$).} \label{abs_low_er}
\end{figure*}

\begin{figure*}
\plotone{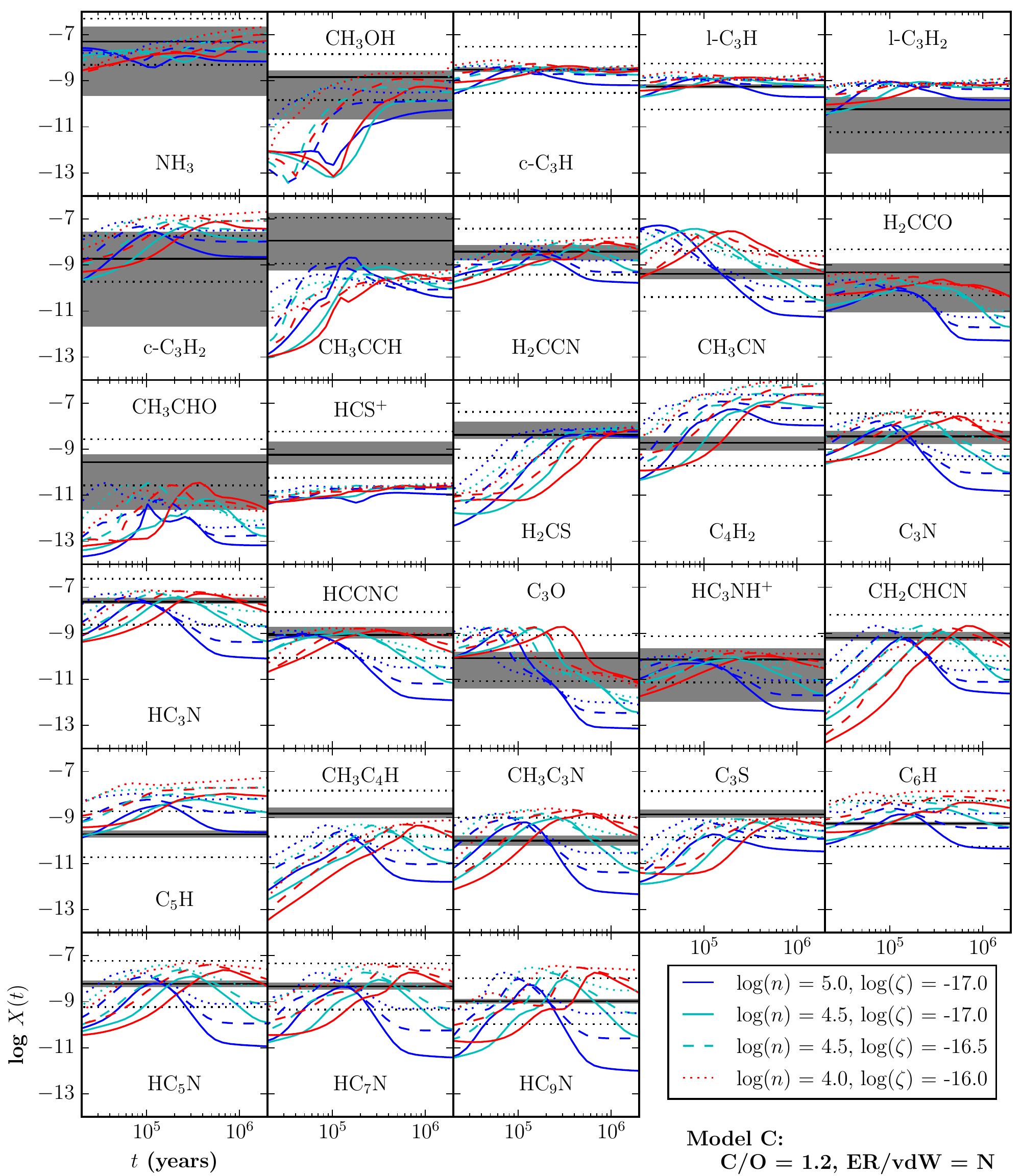}
\caption{Time-dependent abundances log $X(t)$ for Model C (C/O = 1.2, ER/vdW = N, $a$ = 0.01, $b$ = 0.4)
  with observed abundances log $X_{{\rm obs}}$ in solid black, an order of magnitude difference from the observed
  abundances log $X_{{\rm obs}} \pm 1$ in dotted black, and the 1$\sigma$ uncertainties on the observed abundances
  in gray. The linestyle corresponds to the
  cosmic ray ionization rate: solid ($\zeta = 10^{-17}\ $s$^{-1}$), dashed ($\zeta = 10^{-16.5}\ $s$^{-1}$), and dot-dashed
  ($\zeta = 10^{-16}\ $s$^{-1}$), while the color indicates the value of density: blue ($n = 10^5\ {\rm cm}^{-3}$),
  cyan ($n = 10^{4.5}\ {\rm cm}^{-3}$), and red ($n = 10^4\ {\rm cm}^{-3}$).} \label{abs_high}
\end{figure*}

\begin{figure*}
\plotone{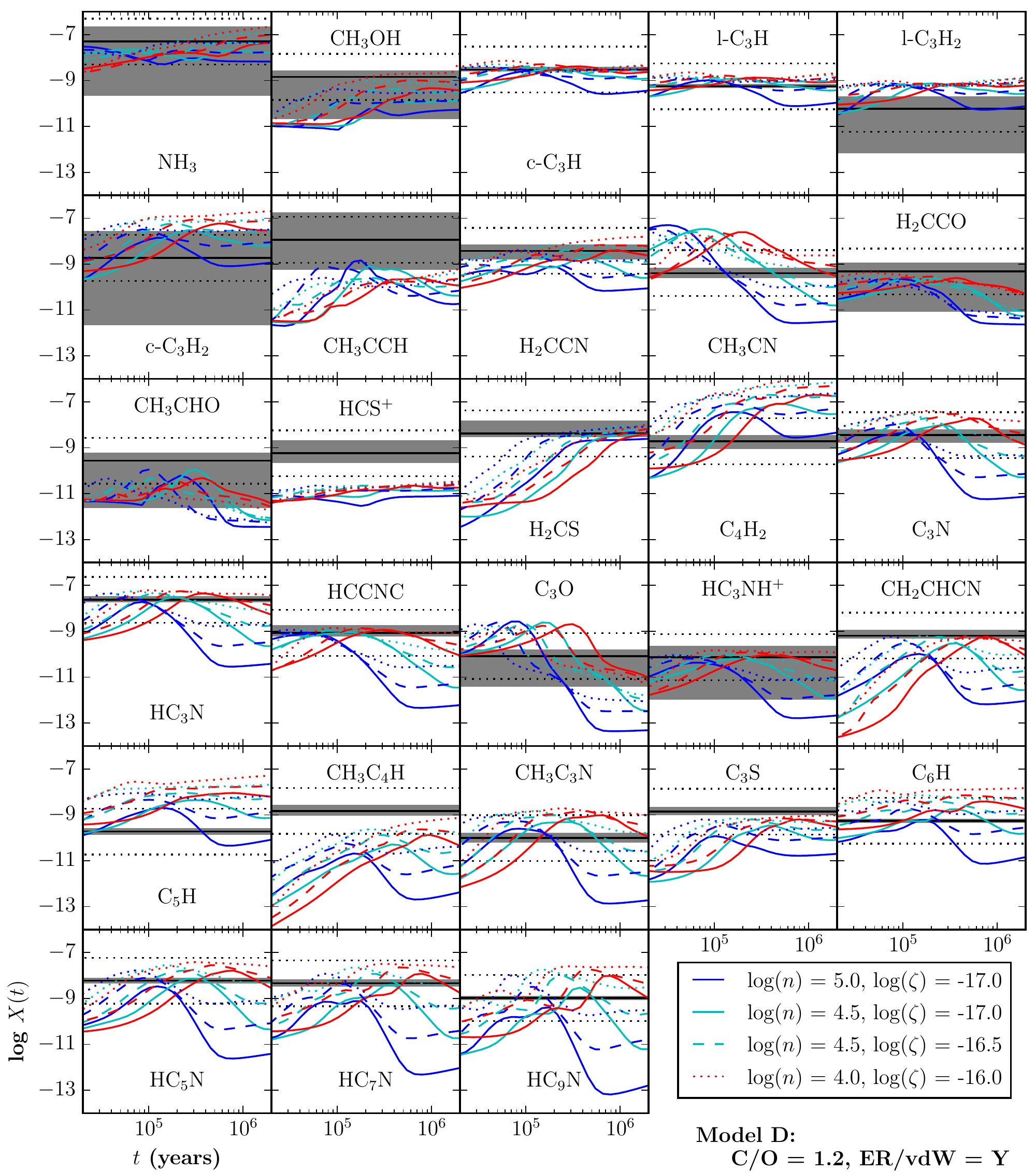}
\caption{Time-dependent abundances log $X(t)$ for Model D (C/O = 1.2, ER/vdW = Y, $a$ = 0.01, $b$ = 0.4)
  with observed abundances log $X_{{\rm obs}}$ in solid black, an order of magnitude difference from the observed
  abundances log $X_{{\rm obs}} \pm 1$ in dotted black, and the 1$\sigma$ uncertainties on the observed abundances
  in gray. The linestyle corresponds to the
  cosmic ray ionization rate: solid ($\zeta = 10^{-17}\ $s$^{-1}$), dashed ($\zeta = 10^{-16.5}\ $s$^{-1}$), and dot-dashed
  ($\zeta = 10^{-16}\ $s$^{-1}$), while the color indicates the value of density: blue ($n = 10^5\ {\rm cm}^{-3}$)
  cyan ($n = 10^{4.5}\ {\rm cm}^{-3}$), and red ($n = 10^4\ {\rm cm}^{-3}$).} \label{abs_high_er}
\end{figure*}

Figures \ref{abs_low}, \ref{abs_low_er}, \ref{abs_high}, and \ref{abs_high_er} show the modeled and observed
time-dependent abundances $X(t)$ of each molecule represented in both the TMC-1 emission line survey analysis
\citep{gratier2016} and the
chemical network (Group G) for Models A, B, C, and D, respectively. Each panel shows the abundances of a
single molecule for different density and ionization rate pairs contained in each restricted
model at single values of the reaction desorption efficiency ($a = 0.01$) and the
diffusion-binding energy ratio ($b = 0.4$). The linestyle corresponds to the cosmic ray ionization rate
(solid for $\zeta = 10^{-17}\ $s$^{-1}$, dashed for $\zeta = 10^{-16.5}\ $s$^{-1}$, and dot-dashed
for $\zeta = 10^{-16}\ $s$^{-1}$), while the color indicates the value of density (blue for $n = 10^5\ {\rm cm}^{-3}$,
cyan for $n = 10^{4.5}\ {\rm cm}^{-3}$, and red for $n = 10^4\ {\rm cm}^{-3}$).

Each time-dependent abundance for each molecule exhibits at least one of four characteristics :

1. Small carbon-containing species
c-C$_3$H, l-C$_3$H, l-C$_3$H$_2$, H$_2$CCN, HCS$^+$ and ammonia NH$_3$ exhibit abundances that show little variation
(\textit{inertness}) between $\sim 10^5$ and $10^6$ years and over the densities and ionization
rates contained in the chemical model parameter space. The abundances of ammonia NH$_3$ and of hydrocarbons c-C$_3$H, l-C$_3$H,
l-C$_3$H$_2$ are within an order of magnitude of the observed values for long periods of time
($t = 2 \times 10^4 - 2 \times 10^6$ years) for all Models A, B, C, and D, while the heavier species
H$_2$CCN, HCS$^+$ are underproduced except in Model C where, as a result of increased oxygen depletion, the relative increase in the elemental carbon abundance
increases the abundances of H$_2$CCN so that it agrees within an order of magnitude of the observed value
for a long period of time.

2. A single peak or trough corresponding to a clear \textit{maximum} or \textit{minimum} abundance is
another common feature. Maxima are present in the oxygen-containing organic
species CH$_3$CHO, the cyanopolyyne HC$_3$N, and the carbon-chain molecules CH$_3$CN,
CH$_3$CCH, H$_2$CCO, C$_4$H$_2$, HCCNC, C$_3$O, HC$_3$NH$^+$, CH$_3$C$_4$H, CH$_3$C$_3$N, C$_3$S.
The carbon-chain C$_6$H presents the only clear minimum between $2 \times 10^4 $ and
$2 \times 10^6$ years in Models A and B. The effect of density and cosmic ray ionization rate on the timescale immediately
appears in each panel where greater densities and greater ionization rates produce earlier timescales
and where lower densities and lower ionization rates effect similar abundance features at later times
so that the timescale appears inversely related to both density and ionization rate.

3. Molecules c-C$_3$H$_2$, CH$_3$OH, H$_2$CS, and C$_4$H$_2$ show a \textit{monotonic}
increase of the time-dependent abundances with no clear maximum or minimum. Two quasi steady-states
(periods of time where the abundances change very little) in the time-dependent abundances of
CH$_3$OH, H$_2$CS, and C$_4$H$_2$ appear at early ($t < 10^5$ year) and late ($t > 10^5$ year) times,
while the abundances of c-C$_3$H$_2$ resemble the inert character with a slight positive
gradient in time.

4. The abundances of cyanopolyynes HC$_5$N, HC$_7$N, and HC$_9$N in addition to the abundances of both
C$_3$N and C$_5$H in Model A \textit{oscillate} in time exhibiting large variation in both magnitude and feature character when
density and cosmic ray ionization rate vary. The abundances of these molecules vary up to five orders of
magnitude, and the oscillatory behavior is exemplified by the larger cyanopolyyne abundances $X_{{\rm HC}_7{\rm N}}(t)$ and
$X_{{\rm HC}_9{\rm N}}(t)$ in Models A and B in which the amplitude of oscillation increases toward low densities and high ionization rates
($n = 10^4$ cm$^{-3}$, $\zeta = 10^{-16}$ s$^{-1}$).

\noindent The characteristic behaviors for the time-dependent abundances of individual molecules
change throughout the grid models as the free parameters vary;
Table \ref{tab_abs} contains a summary of the behavior character types $\mathcal{C}$ = 1, 2, 3, and 4 of the
time-dependent abundances for each Model A, B, C and D in the restricted grid space ($a = 0.01$, $b = 0.4$) shown in Figures \ref{abs_low},
\ref{abs_low_er}, \ref{abs_high}, and \ref{abs_high_er}.

\subsection{Grid Solutions and Agreement}

\begin{figure*}
\plotone{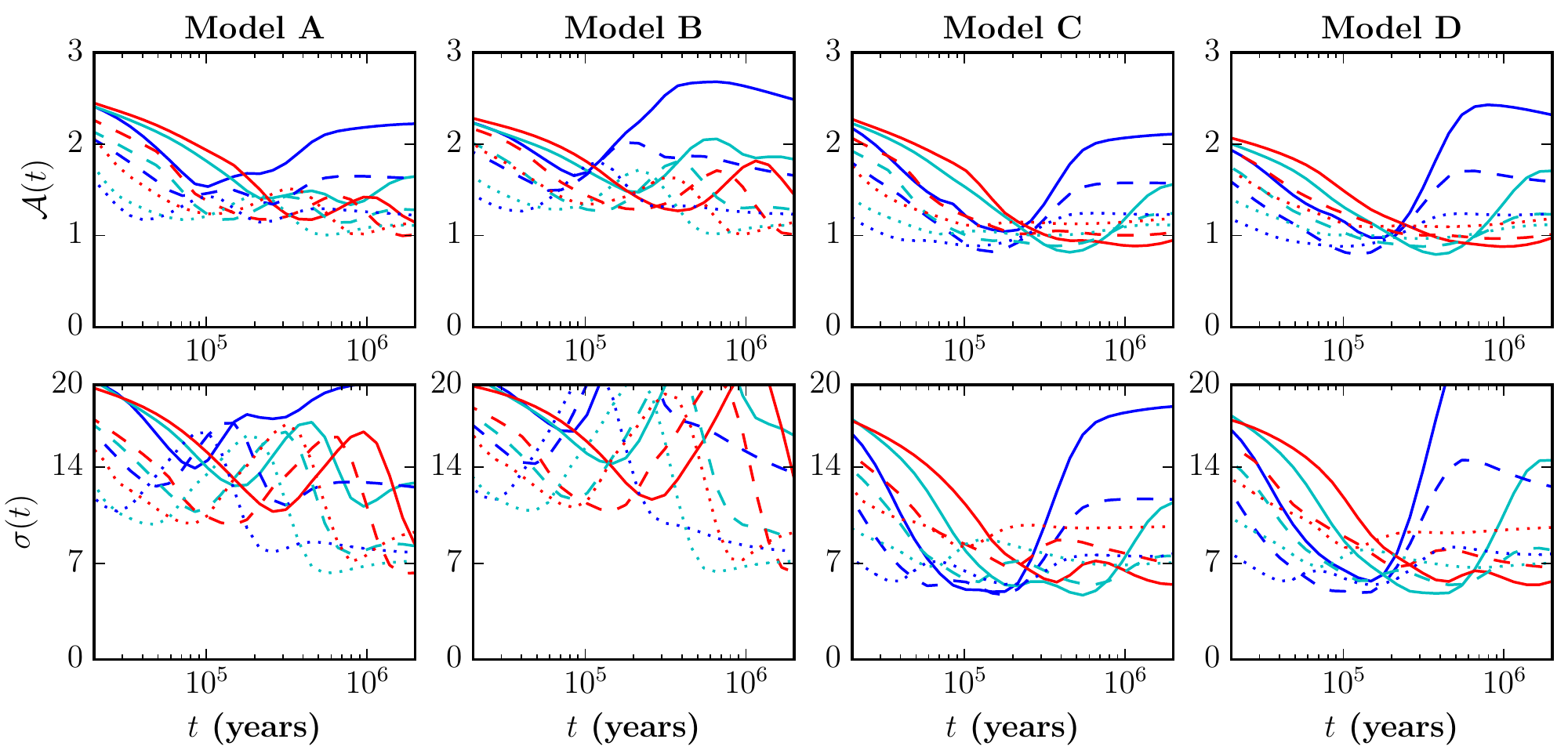}
\caption{Time-dependent rms log abundance differences $\mathcal{A}(t)$ and rms deviations
  $\sigma(t)$ for Group G and Models A, B, C, and D ($a$ = 0.01, $b$ = 0.4). The linestyle corresponds to the
  cosmic ray ionization rate: solid ($\zeta = 10^{-17}\ $s$^{-1}$), dashed ($\zeta = 10^{-16.5}\ $s$^{-1}$), and dot-dashed
  ($\zeta = 10^{-16}\ $s$^{-1}$), while the color indicates the value of density: blue ($n = 10^5\ {\rm cm}^{-3}$),
  cyan ($n = 10^{4.5}\ {\rm cm}^{-3}$), and red ($n = 10^4\ {\rm cm}^{-3}$). \label{fits_g}}
\end{figure*}

\begin{figure*}
\plotone{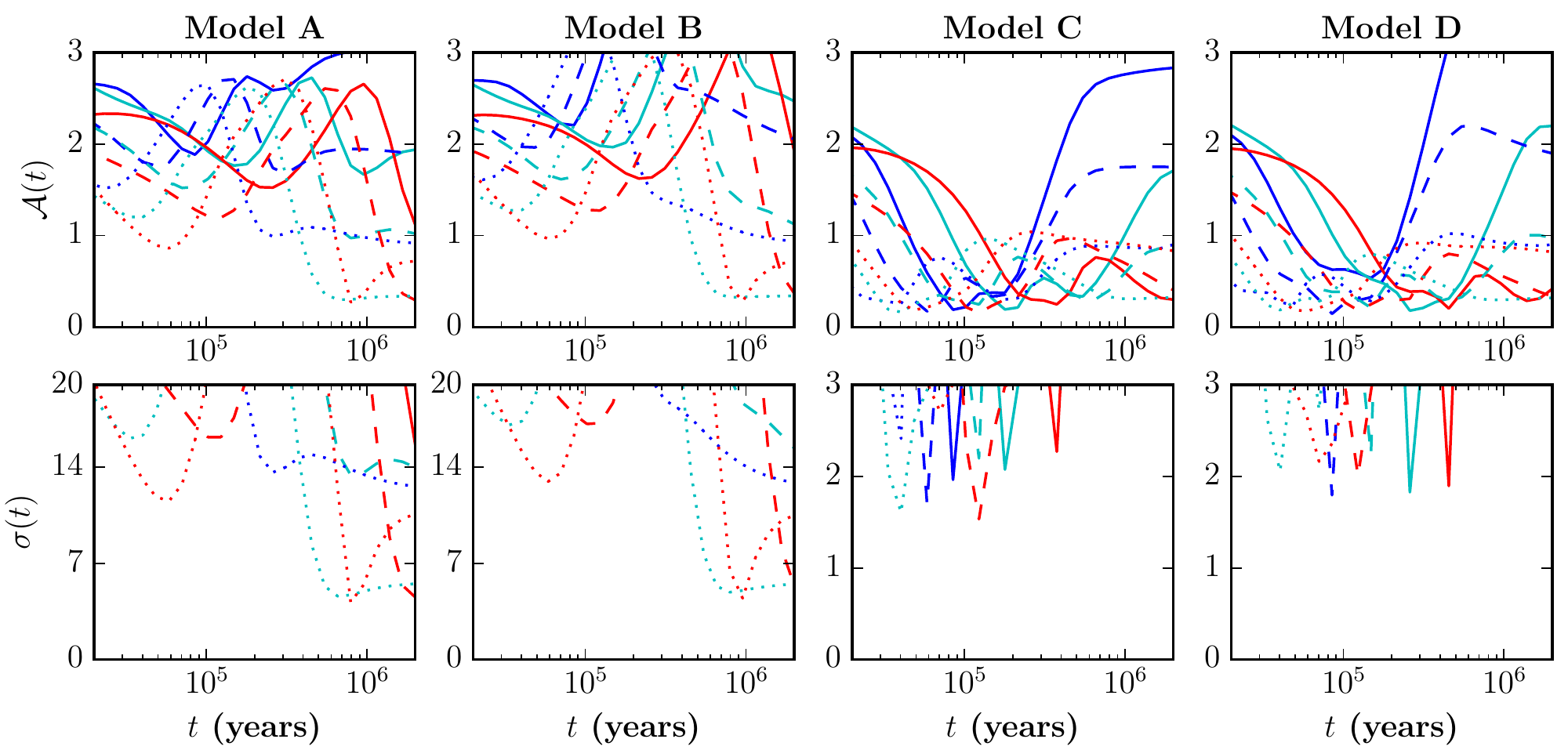}
\caption{Time-dependent rms log abundance differences $\mathcal{A}(t)$ and rms deviations
  $\sigma(t)$ for Group C and Models A, B, C, and D ($a$ = 0.01, $b$ = 0.4). The linestyle corresponds to the
  cosmic ray ionization rate: solid ($\zeta = 10^{-17}\ $s$^{-1}$), dashed ($\zeta = 10^{-16.5}\ $s$^{-1}$), and dot-dashed
  ($\zeta = 10^{-16}\ $s$^{-1}$), while the color indicates the value of density: blue ($n = 10^5\ {\rm cm}^{-3}$),
  cyan ($n = 10^{4.5}\ {\rm cm}^{-3}$), and red ($n = 10^4\ {\rm cm}^{-3}$). \label{fits_c}}
\end{figure*}

Figures \ref{fits_g}, \ref{fits_c}, and \ref{fits_s} show the time-dependent rms log difference 
$\mathcal{A}(t)$ and rms deviation $\sigma(t)$ on appropriate scales over the restricted grids for the
three molecular Groups, G, C, and S, respectively,
and we apply the aforementioned characteristic behavior types to descibe $\mathcal{A}(t)$ and $\sigma(t)$.
For single minima $\mathcal{A}_{{\rm min}}$ with clear concavity, the reasonable span of the chemical timescale depends on the
criterion value $\mathcal{A}_{{\rm crit}}$ describing the region of reasonable agreement $\{t\}$ when $\mathcal{A}_{{\rm min}}(t) < \mathcal{A}_{{\rm crit}}$.
Model fits exhibiting low mean differences or deviations with little variation over extended timescales place weaker constraints on the
solution $\{t\}$ when $\mathcal{A}_{{\rm min}}(t) < \mathcal{A}_{{\rm crit}}$ since many times would satisfy this condition.
When rms differences oscillate, imposing a value of $\mathcal{A}_{{\rm crit}}$ will separate the region of solution
$\mathcal{A}_{{\rm min}}(t) < \mathcal{A}_{{\rm crit}}$ into distinct features when $\mathcal{A}(t) > \mathcal{A}_{{\rm crit}}$ during the
oscillation. Comparing $\mathcal{A}(t)$ and $\sigma(t)$ illustrates the effects the uncertainty type and group sizes
have on the agreement between modeled and observed abundances.

We summarize
$\mathcal{A}_{{\rm min}}$ in Tables \ref{aa}, \ref{ab}, \ref{ac}, and \ref{ad} and $\sigma_{{\rm min}}$ in Tables \ref{sa}, \ref{sb},
\ref{sc}, and \ref{sd} for each point $p$ in the grid  by assigning symbols to illustrate the general agreement of each group
with respect to the observed abundances and uncertainties. We classify the model fit measures $\mathcal{A}_{{\rm min}}$ into three distinct regions:
$\mathcal{A}_{{\rm min}} > 1\ ($open, $ \fullmoon),\ 1 \geq \mathcal{A}_{{\rm min}} \geq .5\ ($dotted,
$ \astrosun),\ \mathcal{A}_{{\rm min}} < 0.5\ ($closed, $ \newmoon)$; the mean deviations follow the same symbolic representation
since the 1$\sigma$ uncertainty values are reported with the observed abundances. We divide the best fit times $\mathcal{T}$ into
four regions: log$(\mathcal{T}) < 5\ ($crescent, $ \leftmoon)$, $5 <$ log$(\mathcal{T}) < 6\ ($open, $ \fullmoon)$,
$6 <$ log$(\mathcal{T}) < 7\ ($dotted, $ \astrosun)$, log$(\mathcal{T}) > 7\ ($closed, $ \newmoon)$ (see the bottom halves of
Tables \ref{aa} - \ref{sd}). The symbolic representation shows the degree and extent of model agreement and consistency
among molecular groups over the grid spaces, and the trends in the solution space characterize the model with respect to the
observed abundances over a large parameter space, naturally benchmarking the chemical network and code in many unique sets of
homogeneous conditions.

To visualize the solution space of our four grid Models A, B, C, and D over four varied free parameters, $n$, $\zeta$, $a$, and $b$,
we record the minimum fit measures for each point in the grid, $\mathcal{A}_{{\rm min}}(p,\mathcal{T})$,
the minimum mean deviations $\sigma_{{\rm min}}(p,\mathcal{T})$, and the corresponding timescales $\mathcal{T}$ for each
grid point $p = \{n,\zeta,a,b\}$ and project the values on inner axes of $n \times \zeta$ and outer axes of $a \times b$.
The remaining parameters $\{T_{{\rm gas}}, A_{{\rm V}}, T_{{\rm dust}}\}$, each of which subtends only a single value, maintain
that single value throughout the analysis and discussion. The symmetry of the minimum fit measure matrices over
the span of densities and ionization rates, $\mathcal{A}_{{\rm min}}^{\zeta \times n}$, emerges across the main diagonal
log($n$) + log($\zeta$) = $-12$ (top left to bottom right, see note in Table \ref{eta_tab}) producing sets of $m$ degenerate solutions
($\mathcal{A}_{{\rm min}}^{1} \sim \mathcal{A}_{{\rm min}}^2 \sim ... \sim \mathcal{A}_{{\rm min}}^m$) along diagonals of
$\eta$ = log($n$) - log($\zeta$) implying $\eta$ is a natural quantification of solutions $\mathcal{A}_{{\rm min}} < 1$ in these
particular grids. For the grids of dark cloud conditions, $\eta$, as seen in Table \ref{eta_tab}, is defined over the
set of half integers between and including the maximum and minimum values of density and ionization rate, $\eta = 22$ when
log($n$) = 5 and log($\zeta$) = $-17$ and $\eta = 20$ when log($n$) = 4 and log($\zeta$) = $-16$, respectively, or
${\rm log}(n) - {\rm log}(\zeta) = \eta = \{20, 20.5, 21, 21.5, 22\}$ and relates to $\zeta/n$ in previous studies by
$\eta = -$log$(\zeta/n)$ \citep{lepp1996,tine1997}.

\subsubsection{The Total Group G}

\begin{deluxetable}{c|ccc}
\tabletypesize{\footnotesize}
\tablecaption{$\eta = {\rm log}(n) - {\rm log}(\zeta)$ \label{eta_tab}}
\tablehead{ log($\zeta$) $\backslash$ log($n$) = & 5 & 4.5 & 4
}
\decimals
\startdata
-17 & 22 & 21.5 & 21 \\
-16.5 & 21.5 & 21 & 20.5 \\
-16 & 21 & 20.5 & 20 \\
\enddata
\tablecomments{The main diagonal satisfies log($n$) + log($\zeta$) = -12}
\end{deluxetable}

Models A and B show little if any solution \{$t$\} when $\mathcal{A}(t) < 1$ is satisfied implying a poor fit with
average factors of agreement greater than an order of magnitude. The elemental composition C/O = 1.2, or the high oxygen depletion case,
improves fits and strengthens the minimum character of $\mathcal{A}(t)$ in Models C and D demonstrating the preference of
Group G abundances to carbon-rich, oxygen-poor conditions. Because
$\mathcal{A}(t) < 1$ is true for short time periods as seen in Figure \ref{fits_g}, each point model solution is only instantaneously
well-constrained in time, and this corresponds to a minimum solution. The oscillatory behavior character of
$\mathcal{A}(t)$ and $\sigma(t)$ in Models A and B changes in Models B and D either to functions with clear minima or to those that decrease monotonically in time.
This change reflects the time-dependent abundances of HC$_7$N and HC$_9$N, which change from oscialltory functions in Models A and B
to functions with clear maxima, larger values, and sharper behavior in Models C and D. The
best instantaneous solution, $\mathcal{A}_{{\rm min}} = 0.66$, corresponding to an average factor of agreement of 4.6 for
the 32 abundances, appears at time $\mathcal{T} =  1.8 \times 10^{5}$ years in Model D for dense conditions
($n = 10^5$ cm$^{-3}$ and $\zeta = 10^{-17}$ s$^{-1}$), a large diffusion-binding energy ratio ($b = 0.5$), and a high reactive
desorption efficiency ($a = 0.1$), though many solutions $\mathcal{A}_{{\rm min}} < 1$ exist throughout the grid models. The small
uncertainties on several of the observed abundances result in large mean deviations $\sigma(t) > 3$ at all times, and the
behavior of $\sigma(t)$ is similar to $\mathcal{A}(t)$, though $\sigma(t)$ appears exaggerated over its larger scale.

The timescales associated with the minimum mean deviations for Group G remain relatively consistent
with those determined by minimizing the rms log difference $\mathcal{A}(p,t)$ between the observed and modeled abundances,
but the range of minimum mean deviations $3.5 < \sigma_{{\rm min}}(p,\mathcal{T}) < 7$ reveals the inability of any single homogeneous model within our grids at any time
to reproduce successfully a set of 32 observed abundances constrained by the uncertainties $\sigma_i$ produced by the
LTE model and Bayesian analysis of the observed emission. Furthermore, because the observed abundance uncertainties vary in
size, the mean deviation does not equivalently quantify the agreement for all species in this group.

\subsubsection{The Cyanopolyyne Group C}

The new observed abundances of the cyanopolyynes (\cite{gratier2016}, see our Table \ref{tab_abs}) differ by only a factor of a
few ($\leq 4$) when compared to previous values \citep{smith2004}, but earlier large gas-grain kinetic models of TMC-1 (CP) using
a low C/O ratio failed to reproduce the cyanopolyyne abundances consistently with the rest of a larger group
\citep{garrod2007}. In our models, the timescales of maximum abundance for HC$_3$N and HC$_5$N vary inversely with both the density and the
ionization rate. Increased oxygen depletion and the increased C/O ratio of Model C moves the maximum abundances of HC$_3$N, HC$_5$N, and HC$_7$N of the darkest model
$n = 10^5$ cm$^{-3}$ and $\zeta = 10^{-17}$ s$^{-1}$ very close to the observed values determined from the TMC-1 (CP) emission, 
and an instantaneous quasi-steady state emerges at $10^5$ years for these three cyanopolyyne abundances. The abundance of
HC$_7$N exhibits the same behavior as the abundances of the smaller two cyanopolyynes over the restricted grid space $(b = 0.4$ and $a = 0.01$)
of Models C and D but oscillates in time in Models A and B. The maximum abundances over the restricted grid space for the first three cyanopolyynes remain
within an order of magnitude of the observed values for extended periods throughout Models C and D. The final
observed cyanopolyyne, HC$_9$N, exhibits underproduction in Model A, and while the increased C/O ratio of Model C generally
induces a shift of the abundances of the restricted grid space to within an order of magnitude of the observed value for
extended timescales, significant overproduction of HC$_9$N skews the average agreement in Model C, where this overproduction
increases the modeled abundances beyond an order of magnitiude above the observed value.
Furthermore, in contrast to the smaller cyanopolyynes, the abundance of HC$_9$N in the darkest model $n = 10^5$ cm$^{-3}$ and $\zeta = 10^{-17}$ s$^{-1}$ in
Model C is no longer as well fit to the observed abundance.

The cyanopolyynes are formed in the gas phase via the dissociative recombination of protonated precursors:
\begin{equation}
 {\rm H}{\rm C}_{2n+1}{\rm NH}^+ + {\rm e}^- \rightarrow {\rm HC}_{2n+1}{\rm N} + {\rm H}
\end{equation}
\noindent and from reactions
\begin{equation}
  {\rm C}_{2n+2}{\rm H} + {\rm N} \rightarrow  {\rm HC}_{2n+1}{\rm N} + {\rm C} \label{eqn}
\end{equation}
\noindent and
\begin{equation}
 {\rm C}_{2n}{\rm H}_2 + {\rm CN} \rightarrow  {\rm HC}_{2n+1}{\rm N} + {\rm H}
\end{equation}
\noindent between atomic and neutral radicals and other carbon-chain molecules lacking nitrogen.
A recent emission line survey of cyanopolyyne carbon isotopologues toward TMC-1
\citep{burkhardt2017} showed by process of elimination that the reactions between hydrocarbon ions and
atomic nitrogen must dominate the production of cyanoppolyynes HC$_5$N and HC$_7$N, but the proposed chemical network lacks
reactions between neutral carbon chains and atomic nitrogen similar to equation (\ref{eqn}) as a
possible mechanism of formation. The cyanopolyynes are destroyed through ion-neutral reactions in the gas
with abundant ions that produce the protonated precursors and from neutral-neutral reactions of the type
\begin{equation}
  {\rm H}{\rm C}_{2n+1}{\rm N} + {\rm C} \rightarrow {\rm C}_{2n+2}{\rm N} + {\rm H}
\end{equation}
\noindent with abundant atomic species such as carbon.

\begin{figure*}
\plotone{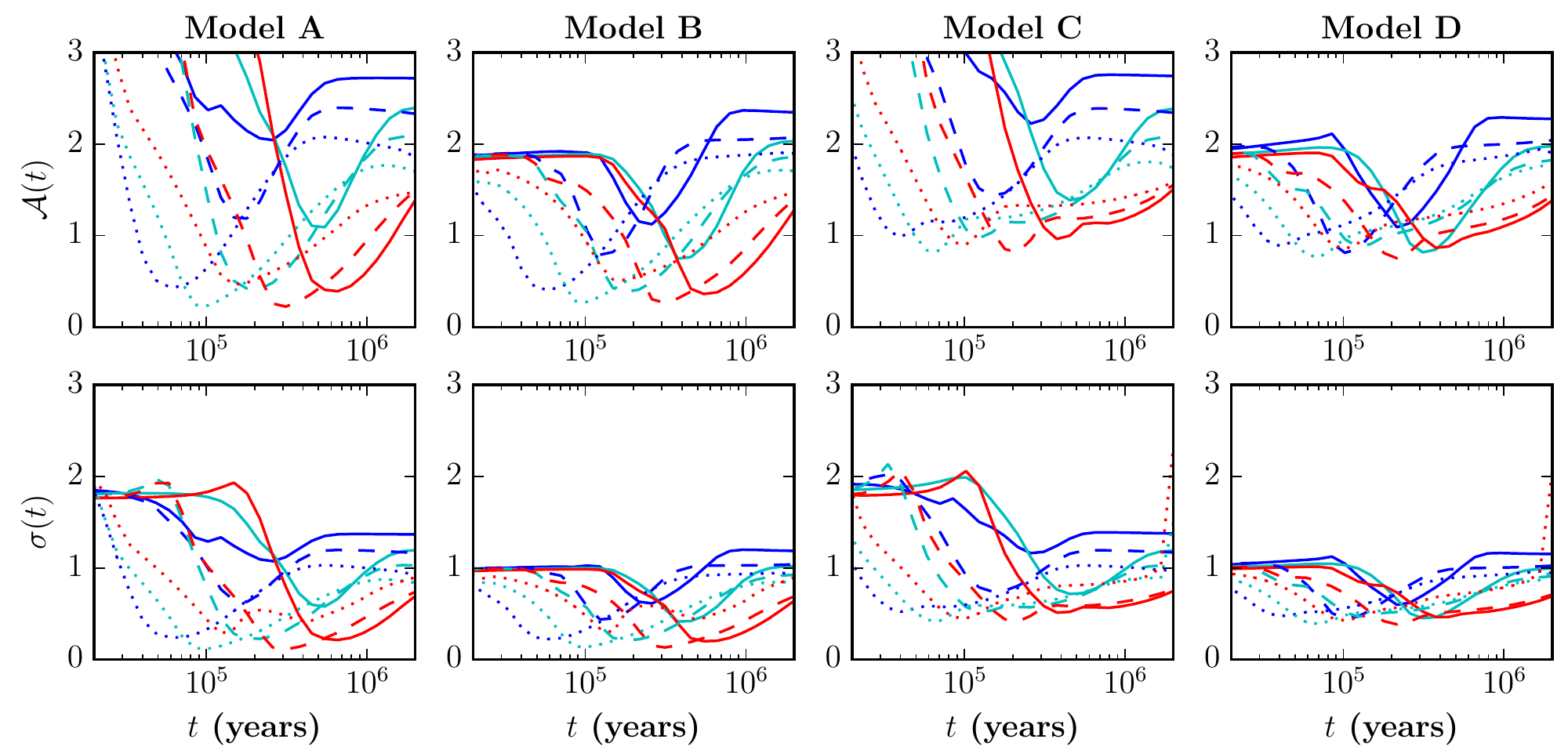}
\caption{Time-dependent rms log abundance differences $\mathcal{A}(t)$ and rms deviations
  $\sigma(t)$ for Group S and Models A, B, C, and D ($a$ = 0.01, $b$ = 0.4). The linestyle corresponds to the
  cosmic ray ionization rate: solid ($\zeta = 10^{-17}\ $s$^{-1}$), dashed ($\zeta = 10^{-16.5}\ $s$^{-1}$), and dot-dashed
  ($\zeta = 10^{-16}\ $s$^{-1}$), while the color indicates the value of density: blue ($n = 10^5\ {\rm cm}^{-3}$),
  cyan ($n = 10^{4.5}\ {\rm cm}^{-3}$), and red ($n = 10^4\ {\rm cm}^{-3}$).\label{fits_s}}
\end{figure*}

The dissociative recombination of the protonated precursors dominates the production of all of the cyanopolyynes
at the best fit times along the restricted main diagonal except for the two larger cyanopolyynes, HC$_7$N and
HC$_9$N, in the darkest conditions, $\eta = 22$, and the least dark conditions, $\eta = 20$, where the
neutral-neutral reactions involving atomic nitrogen and the cyano radical contribute most to their production,
respectively, at the chemical timescales. Increasing both oxygen depletion and the carbon-oxygen
ratio in Model C to C/O = 1.2 shifts the dominant production pathways of all of the cyanopolyynes to
neutral-neutral reactions at the best times for all
$\eta$ along the restricted main diagonal, and substantial destruction involving the cyano radical also emerges as a
results of the elemental composition.

Characteristic of the abundances of the two large cyanopolyynes $X_{{\rm HC}_7{\rm N}}(t)$ and $X_{{\rm HC}_9{\rm N}}(t)$, the
rms log difference $\mathcal{A}(t)$ for Group C exhibits oscillatory behavior
with respect to time in Models A and B where solutions $\mathcal{A}(t) < 1$ appear at long timescales
($t > 5 \times 10^5$ years) and improve with respect to decreasing density and
increasing ionization rate. The amplitude of the oscillations increases as the Eley-Rideal and van der
Waals complex reaction mechanisms are activated, but the overall behavior of $\mathcal{A}(t)$
with respect to the models without the new mechanisms is preserved. The small uncertainties in
the cyanopolyyne abundances result in large mean deviations ($\sigma(t) > 3$) for all models in grid Models A and B
and can be seen in Figure \ref{fits_c}. A very large solution region with $\mathcal{A}(t) < 1$
emerges in our restricted grid Models C and D with increased oxygen depletion for Group C, and the mean deviations
illustrate the exaggeration of the behavior of $\mathcal{A}(t)$ as $\sigma(t)$ sharply changes
around its minima. The best solution for Group C, $\mathcal{A}_{{\rm min}} = 0.14$, corresponding to
an average factor of agreeement of 1.4, appears in Model C at a slightly earlier time
($\mathcal{T} = 7.1 \times 10^4$ years) than Group G but with the same dense conditions
($n = 10^5$ cm$^{-3}$ and $\zeta = 10^{-17}$ s$^{-1}$), large diffusion-binding energy ratio ($b = 0.5$), and high
reactive desorption efficiency ($a = 0.1$).

Similar to Group G, the timescales associated with the minimum deviation of Group C are very similar
to those determined using the unweighted fit method. In contrast to the generally good agreement reflected by
$\mathcal{A}_{{\rm min}}(p, \mathcal{T})$, the range of minimum rms deviations,
$1.5 < \sigma_{{\rm min}}(p, \mathcal{T}) < 4.5$, reflects the constraining effect several small uncertainties have
on the agreement between the chemical kinetic model and the abundances derived from an LTE model of the molecular emission spectra observed toward TMC-1 (CP).
Although our models fit the observed
abundances often within a factor of a few, we cannot reproduce at any time the cyanopolyyne abundances within a mean
standard deviation of the abundances constrained by the prior distributions of observed column densities
determined from the Bayesian anaylsis of the LTE model of the numerous observed emission lines.

\subsubsection{The Oxygen-containing Organic Species Group S}

\input moon_tmc1_low_rev.tex
\input moon_tmc1_low_er_rev.tex 
\input moon_tmc1_high_rev.tex
\input moon_tmc1_high_er_rev.tex

\input moon_sigma_tmc1_low_rev.tex
\input moon_sigma_tmc1_low_er_rev.tex 
\input moon_sigma_tmc1_high_rev.tex
\input moon_sigma_tmc1_high_er_rev.tex

Initial models of TMC-1 implementing reactive desorption \citep{garrod2007} showed enhancement of the abundance of gas-phase
acetaldehyde over two orders of magnitude, but even with high reactive desorption efficiency  ($a = 0.1$),
the acetaldehyde abundance $X_{{\rm mod}}({\rm CH}_3{\rm CHO}) = 1.7 \times 10^{-11}$ did not sufficiently reproduce the observed
value $X_{{\rm obs}}({\rm CH}_3{\rm CHO}) = 6 \times 10^{-10}$.
In contrast, these first models sufficiently reproduced the observed abundance of gas-phase methanol
$X_{{\rm obs}}({\rm CH}_3{\rm OH}) = 3 \times 10^{-9}$  at the best fit time for models with moderate reactive desorption
efficiency: $X_{{\rm mod}}({\rm CH}_3{\rm OH}) = 1.1 \times 10^{-9}$ for $a = 0.03$.

The gas-phase abundance of methanol challenged astrochemical kinetic models until nonthermal desorption
enabled some of the methanol formed via exothermic surface reactions to return to the gas upon formation
following the hydrogenation of lighter species on the grain surfaces,
\begin{equation}
 s-{\rm CH}_3{\rm O} + s-{\rm H} \rightarrow {\rm CH}_3{\rm OH}
  \end{equation}
\noindent and
\begin{equation}
 s-{\rm CH}_2{\rm OH} + s-{\rm H} \rightarrow {\rm CH}_3{\rm OH},
\end{equation}
\noindent and this mechanism dominates the production at all times $\mathcal{T}$ along the main diagonal
$\eta$ = 22 ($n = 10^5$ cm$^{-3}$ and $\zeta = 10^{-17}$ s$^{-1}$), 21 ($n = 10^{4.5}$ cm$^{-3}$ and $\zeta = 10^{-16.5}$ s$^{-1}$),
and 20 ($n = 10^4$ cm$^{-3}$ and $\zeta = 10^{-16}$ s$^{-1}$). In Model B, fragmentation through dissociative recombination of protonated
dimethyl ether
\begin{equation}
 {\rm CH}_3{\rm OCH}_4^+ + {\rm e}^- \rightarrow {\rm CH}_3{\rm OH} + {\rm CH}_3
\end{equation}
\noindent dominates the production in the dark model $\eta = 22$ at time $\mathcal{T} = 2.6 \times 10^5$ years
as a result of the additional ice-surface dimethyl ether created from
the successive hydrogenation of a carbon-methanol van der Waals complex.
Even with low nonthermal desorption efficiency ($a = 0.01$),
the gas phase abundance of dimethyl ether is enhanced by several orders of magnitude
at all times $t < 3 \times 10^7$ years as a result of the increased ice abundance. The abundance of protonated dimethyl ether, which forms in the gas through the reaction between gas-phase dimethyl
ether and abundant ions,
successfully fuels the gas-phase dissociative recombination reaction to competitive levels with respect to the other methanol
formation mechanisms of Model A. Methanol is destroyed by ion-neutral and neutral-neutral
reactions with abundant ions and atoms in the gas, respectively. 

Acetaldehyde exhibits the reverse situation: in Model A without the new mechanisms, the neutral-neutral reactions
\begin{equation}
 {\rm C}_3{\rm H}_7 + {\rm O} \rightarrow {\rm CH}_3{\rm CHO} + {\rm CH}_3
\end{equation}
\noindent and
\begin{equation}
 {\rm C}_2{\rm H}_5 + {\rm O} \rightarrow {\rm CH}_3{\rm CHO} + {\rm H}
\end{equation}
\noindent between hydrocarbons and atomic oxygen in the gas primarily constitute the overall production along the restricted main diagonals with slight
contribution from the dissociative recombination reactions
\begin{equation}
 {\rm C}_2{\rm H}_5{\rm OH}_2^+ + {\rm e}^- \rightarrow {\rm CH}_3{\rm CHO} + {\rm H}_2 + {\rm H}
\end{equation}
\noindent and
\begin{equation}
 {\rm CH}_3{\rm CHOH}^+ + {\rm e}^- \rightarrow {\rm CH}_{3}{\rm CHO} + {\rm H}
\end{equation}
\noindent of larger protonated precursors.
When the new mechanisms are activated in Model B, nonthermal desorption via the hydrogenation of surface-bound
acetyl radical
\begin{equation}
 s-{\rm CH}_3{\rm CO} + {\rm H} \rightarrow {\rm CH}_{3}{\rm CHO}.
\end{equation}
\noindent comes to dominate the gas-phase acetaldehyde production in the darkest model ($\eta = 22$).
Similar to the other species, acetaldehyde is destroyed in the gas upon reacting with abundant ionic and neutral carbon,
and the abundances for methanol and acetaldehyde, over the restricted grid spaces of Models A and B, show the same dependence on
$\zeta$ and $n$ as well as the enhancement from the new mechanisms.

The rms log differences $\mathcal{A}(t)$ and deviations $\sigma(t)$ for Group S all appear as distinct minima,
with better fits for low densities and high ionization rates. The new mechanisms improve the fit of the densest model
($n = 10^5\ {\rm cm}^{-3}$ and $\zeta = 10^{-17}\ $s$^{-1}$) in Model B, and the greater oxygen abundance of Models A and B results in a
minimum solution $\mathcal{A}(t) < 1$ over larger regions of the grids. In contrast to the total group and the cyanopolyynes, the character of
the rms deviation $\sigma(t)$ of Group S appears relatively flat in time with $\sigma(t) < 1 $ for long time
periods in all but the densest conditions, where a minimum fit is seen in Figure \ref{fits_s}. In oxygen-rich conditions,
the combination of new mechanisms and large lower uncertainties on the observed abundance values results in simultaneous good
agreement $\mathcal{A}(t) < 1$ and high confidence $\sigma(t) < 1$ for the oxygen-containing organic species in Group S.

Model C contains the best average factor of agreement of 1.06 at time $\mathcal{T} = 3.1 \times 10^5$ years at low density
($n = 10^4$ cm$^{-3}$ and $\zeta = 10^{-17}$ s$^{-1}$), small diffusion-binding energy ratio ($b = 0.3$), and high reactive desorption
efficiency ($a = 0.1$). All Models A, B, C, and D have best average factors of agreement < 1.2, and the abundances of
methanol and acetaldehyde are well-fit simultaneously in large regions of the grids.
At time $\mathcal{T} = 1.5 \times 10^5$ years, the lowest average deviation $\sigma_{{\rm min}} = 0.05$ appears in Model B with
similar parameters to the best average factor of agreement with a slightly faster cosmic ray ionization rate. The low
diffusion-binding energy ratio ($b = 0.3$) of these best fits agrees with the best value in previous three-phase chemical models of
methanol \citep{ruaud2016}.

\subsection{Effect of New Mechanisms and Elemental Composition}

Many of the carbon-chain molecules including the cyanopolyynes HC$_5$N, HC$_7$N, and HC$_9$N are underproduced in Model A and
oxygen-rich conditions (C/O = 0.7) and appear to oscillate over the restricted grid space ($a = 0.01$ and $b = 0.4$).
Activating the Eley-Rideal and van der Waals complex reaction mechanisms (Model B) results in lowered abundances at later times
($t > 10^5$ year) for all cyanopolyynes HC$_3$N, HC$_5$N, HC$_7$N, and HC$_9$N and many other carbon-containing species, though
this effect generally does not result in significant movement from the region of solution for the observed abundances.
The two oxygen-containing organic species CH$_3$OH and CH$_3$CHO benefit from the additional production from the new
mechanisms and increase in abundance at early times creating periods between $2 \times 10^4 $
and $2 \times 10^6$ years within the lower uncertainty limit in the restricted grids. Increasing the elemental oxygen depletion in
Model C (C/O = 1.2) greatly enhances the abundances of many of the carbon-containing molecules leading to better fits
for H$_2$CCN, HC$_3$N, HCCNC, HC$_3$NH$^+$, CH$_2$CHCN, CH$_3$C$_4$H, CH$_3$C$_3$N, C$_3$S, C$_6$H, HC$_3$N, HC$_5$N, HC$_7$N,
and HC$_9$N, but the oxygen-containing species CH$_3$CHO and H$_2$CCO suffer decreased peak abundances in this
elemental composition. Model D reveals the composite effect of oxygen-poor conditions and active new mechanisms on the modeled
abundances of the carbon-chain molecules, which remain good fits to the observed values despite the detrimental effect of the
new mechanisms. Similarly, the modeled abundances for oxygen-containing organic species CH$_3$OH and CH$_3$CHO remain
good solutions with increased oxygen depletion when the new mechanisms are active.

\subsection{Effect of Reactive Desorption and the Diffusion-binding Energy Ratio}

Moderate to high values $a = 0.03$ and $0.1$ of the reactive desorption efficiency produce improved fits in Models A and B
for all Groups G, C, and S where solutions following the criterion $\mathcal{A}_{{\rm min}} < 1$ emerge for less dense models
$\eta \leq 21$. An increase in oxygen depletion (C/O = 1.2) in Models C and D results in overall better fits for Groups G and C
at lower reactive desorption efficiencies ($a = 0.01$) while Group S exhibits weaker fits that can be mitigated with increasing
reactive desorption efficiency ($a = 0.03, 0.1$). The diffusion-binding energy ratio $b$ has a marginal effect on
$\mathcal{A}_{{\rm min}}$, which is fairly constant with respect to $b$ for every group, implying that the
best fits and times are not sensitive to this parameter over the grid spaces. Though individual molecules may present
large sensitivities to $b$, each group fit is not significantly impacted by this parameter alone. 

\section{Summary}

An astrochemical kinetic grid model is an array of time-dependent abundances $X_i(t)$ for molecules $i$ in a chemical network calculated
for different physical conditions and over ranges of free parameters in the chemical model. To solve the rate law equations and obtain the
time-dependent abundances for the physical conditions determined
from the emission of tracers of both compact, dense material (cyanoacetylene) and an extended region of emission (methanol) along
the line of sight to TMC-1 (CP), we parallelize the execution of the rate solver over the span of the representative grid space
for dark cloud conditions. For each molecule $i$ in the chemical network, we construct a 9-D grid of abundances,
\{$X_i,t,T_{gas},n,A_{{\rm V}},\zeta,T_{dust},a,b$\} and attempt to account for chemical and physical
heterogeneity along a single line of sight by grouping observed abundances according to similar chemistry and
minimizing a measure of the differences between the observed and modeled abundance values. The rms log
difference $\mathcal{A}$ between the modeled and observed abundances parametrizes the average factor of agreement, while the
rms deviation $\sigma$ quantifies
the average factor of agreement only when the uncertainties in the group resemble each other; we minimize both of
these for all $points$ $p$ in the grid corresponding to all combinations of free parameters $\{n, \zeta, a, b\}$ and record the resultant timescales.
We compare each method by tabulating $\mathcal{A}_{{\rm min}}(\mathcal{T})$ and $\sigma_{{\rm min}}(\mathcal{T})$
for each of the $points$ in each grid, and we juxtapose values for three different groups of molecules to show how reducing a
large group of observed molecular abundances to smaller groups with chemical similarity resolves, for the sets of observed abundance values,
a solution space throughout the grid models that generally exhibits better agreement than that of the large group. Grid models and
an extensive parameter space allow for competing effects in the model to be separated revealing the unique solution profiles for
each group.

Some salient features of our calculations are listed below:

$\bullet$ The fits of all Groups G, C, and S are sensitive to the density $n$, cosmic ray ionization rate $\zeta$,
the reactive desorption efficiency $a$, and the carbon-oxygen elemental abundance ratio, C/O.

$\bullet$ The Eley-Rideal and van der Waals reaction mechanisms enhance the production of the oxygen-containing organic species leading
to better fits over the model space.

$\bullet$ The diffusion-binding energy ratio $b$ effects marginal changes in the agreement between our models and the observed
abundances toward TMC-1 (CP).

$\bullet$ The oxygen-containing organic species of Group S prefer a low carbon-oxygen elemental abundance ratio (C/O = 0.7) in
contrast to the composite Group G and the cyanopolyynes of Group C, both of which show better agreement with TMC-1 (CP) abundances
for models utilizing a higher carbon-oxygen ratio (C/O = 1.2).

$\bullet$ The solution space of the cyanopolyynes, Group C, extends to include dense models consistent with the observed
abundances in high oxygen depletion conditions, though the dominant chemical pathways at the best fit times shift from dissociative recombination of
the protonated precursors to neutral-neutral pathways in
the dark conditions given by $\eta = 22$.

$\bullet$ Small observed uncertainties for Group C result in no solution $\sigma_{{\rm min}} <1$
in any model, and this is similarly seen for Group G.

$\bullet$ The character of the time-dependent rms log difference between observed and modeled abundances is
preserved, albeit scaled, when calculating the rms deviation, and the quality of agreement reflected by the rms deviation
depends on the uncertainties within the group.

\noindent As interferometric maps of molecular emission become more widely available and as the physical conditions of each
molecular component are determined independently along a single line of sight, grids of chemical models will continue to provide
a multidimensional approach to mapping chemical structure and will check the consistency among models of molecular emission
and the chemical networks used to express the time-dependent chemical structure of groups of molecular components with coupled chemistries.

\acknowledgments

E.H. thanks the National Science Foundation for its support of the astrochemistry program
at the University of Virginia through grant AST 1514884. T.V.W. is supported by the NSF through
the Grote Reber Fellowship Program administered by Associated
Universities, Inc./National Radio Astronomy Observatory, the D.N. Batten Foundation Fellowship from
the Jefferson Scholars Foundation, the Mars Foundation Fellowship from the Achievement Rewards for
College Scientists Foundation, and the Virginia Space Grant Consortium.

\software{Nautilus \citep{ruaud2016}}

\end{document}

%% file: moon_tmc1_low_rev.tex
\begin{deluxetable*}{cc|ccc|ccc|ccc}
\tabletypesize{\tiny}
\tablecaption{$\mathcal{A}_{{\rm min}}(\mathcal{T})$ and log($\mathcal{T}$) for Groups G, C, and S; Model A \label{aa}}
\tablehead{ & $a$ = & 0.01 & 0.01 & 0.01 & 0.03 & 0.03 & 0.03 & 0.10 & 0.10 & 0.10 \\
$db$ & log($\zeta$) $\backslash$ log($n$) = & 5 & 4.5 & 4 & 5 & 4.5 & 4 & 5 & 4.5 & 4
}
\decimals
\startdata
0.3 & -17 & \fullmoon\fullmoon\fullmoon & \fullmoon\fullmoon\fullmoon & \fullmoon\astrosun\newmoon & \fullmoon\fullmoon\fullmoon & \fullmoon\fullmoon\newmoon & \astrosun\newmoon\newmoon & \fullmoon\fullmoon\astrosun & \astrosun\astrosun\newmoon & \astrosun\newmoon\newmoon\\
0.3 &-16.5 & \fullmoon\fullmoon\fullmoon & \fullmoon\astrosun\newmoon & \fullmoon\newmoon\newmoon & \fullmoon\fullmoon\newmoon & \astrosun\newmoon\newmoon & \astrosun\newmoon\newmoon & \astrosun\astrosun\newmoon & \astrosun\newmoon\newmoon & \astrosun\newmoon\newmoon\\
0.3 & -16 & \fullmoon\astrosun\newmoon & \fullmoon\newmoon\newmoon & \fullmoon\newmoon\newmoon & \astrosun\astrosun\newmoon & \astrosun\newmoon\newmoon & \astrosun\newmoon\newmoon & \astrosun\newmoon\newmoon & \astrosun\newmoon\newmoon & \astrosun\newmoon\newmoon\\
\hline
0.4 & 17 & \fullmoon\fullmoon\fullmoon & \fullmoon\fullmoon\fullmoon & \fullmoon\astrosun\newmoon & \fullmoon\fullmoon\fullmoon & \fullmoon\fullmoon\newmoon & \astrosun\newmoon\newmoon & \fullmoon\fullmoon\astrosun & \astrosun\astrosun\newmoon & \astrosun\newmoon\newmoon\\
0.4 & -16.5 & \fullmoon\fullmoon\fullmoon & \fullmoon\astrosun\newmoon & \astrosun\newmoon\newmoon & \fullmoon\fullmoon\newmoon & \astrosun\newmoon\newmoon & \astrosun\newmoon\newmoon & \astrosun\astrosun\newmoon & \astrosun\newmoon\newmoon & \astrosun\newmoon\newmoon\\
0.4 & -16 & \fullmoon\astrosun\newmoon & \fullmoon\newmoon\newmoon & \fullmoon\newmoon\newmoon & \astrosun\astrosun\newmoon & \astrosun\newmoon\newmoon & \astrosun\newmoon\newmoon & \astrosun\newmoon\newmoon & \astrosun\newmoon\newmoon & \astrosun\newmoon\newmoon\\
\hline
0.5 &-17 & \fullmoon\fullmoon\fullmoon & \fullmoon\fullmoon\fullmoon & \fullmoon\astrosun\newmoon & \fullmoon\fullmoon\fullmoon & \fullmoon\fullmoon\newmoon & \astrosun\newmoon\newmoon & \fullmoon\fullmoon\astrosun & \astrosun\astrosun\newmoon & \astrosun\newmoon\newmoon\\
0.5 & -16.5& \fullmoon\fullmoon\fullmoon & \fullmoon\astrosun\newmoon & \astrosun\newmoon\newmoon & \fullmoon\fullmoon\newmoon & \astrosun\newmoon\newmoon & \astrosun\newmoon\newmoon & \astrosun\astrosun\newmoon & \astrosun\newmoon\newmoon & \astrosun\newmoon\newmoon\\
0.5 &-16& \fullmoon\astrosun\newmoon & \astrosun\newmoon\newmoon & \fullmoon\newmoon\newmoon & \astrosun\astrosun\newmoon & \astrosun\newmoon\newmoon & \astrosun\newmoon\newmoon & \astrosun\newmoon\newmoon & \astrosun\newmoon\newmoon & \astrosun\newmoon\newmoon\\
\hline
\hline
0.3 & -17 & \fullmoon\leftmoon\fullmoon & \fullmoon\newmoon\fullmoon & \newmoon\newmoon\fullmoon & \fullmoon\leftmoon\fullmoon & \newmoon\newmoon\fullmoon & \newmoon\newmoon\fullmoon & \leftmoon\leftmoon\fullmoon & \newmoon\newmoon\fullmoon & \newmoon\newmoon\fullmoon\\
0.3 &-16.5 & \leftmoon\astrosun\fullmoon & \astrosun\astrosun\fullmoon & \astrosun\astrosun\fullmoon & \astrosun\astrosun\fullmoon & \astrosun\astrosun\fullmoon & \astrosun\astrosun\fullmoon & \astrosun\astrosun\fullmoon & \astrosun\astrosun\fullmoon & \astrosun\newmoon\fullmoon\\
0.3 & -16 & \astrosun\astrosun\leftmoon & \fullmoon\fullmoon\fullmoon & \fullmoon\astrosun\fullmoon & \astrosun\astrosun\leftmoon & \fullmoon\fullmoon\leftmoon & \fullmoon\astrosun\fullmoon & \astrosun\astrosun\leftmoon & \fullmoon\astrosun\leftmoon & \fullmoon\newmoon\leftmoon\\
\hline
0.4 & 17 & \fullmoon\leftmoon\fullmoon & \fullmoon\fullmoon\fullmoon & \astrosun\newmoon\fullmoon & \fullmoon\leftmoon\fullmoon & \newmoon\newmoon\fullmoon & \newmoon\newmoon\fullmoon & \fullmoon\leftmoon\fullmoon & \newmoon\newmoon\fullmoon & \newmoon\newmoon\fullmoon\\
0.4 & -16.5 & \fullmoon\fullmoon\fullmoon & \fullmoon\astrosun\fullmoon & \astrosun\astrosun\fullmoon & \astrosun\astrosun\fullmoon & \astrosun\astrosun\fullmoon & \astrosun\astrosun\fullmoon & \astrosun\astrosun\fullmoon & \astrosun\astrosun\fullmoon & \astrosun\astrosun\fullmoon\\
0.4 & -16 & \fullmoon\astrosun\leftmoon & \fullmoon\fullmoon\fullmoon & \fullmoon\fullmoon\fullmoon & \astrosun\astrosun\leftmoon & \fullmoon\fullmoon\leftmoon & \fullmoon\astrosun\fullmoon & \astrosun\astrosun\leftmoon & \fullmoon\fullmoon\leftmoon & \fullmoon\newmoon\leftmoon\\
\hline
0.5 &-17 & \fullmoon\leftmoon\fullmoon & \fullmoon\fullmoon\fullmoon & \astrosun\newmoon\fullmoon & \fullmoon\leftmoon\newmoon & \fullmoon\newmoon\fullmoon & \newmoon\newmoon\fullmoon & \leftmoon\leftmoon\fullmoon & \newmoon\newmoon\fullmoon & \newmoon\newmoon\fullmoon\\
0.5 & -16.5& \fullmoon\fullmoon\fullmoon & \fullmoon\astrosun\fullmoon & \astrosun\astrosun\fullmoon & \fullmoon\astrosun\fullmoon & \astrosun\astrosun\fullmoon & \astrosun\astrosun\fullmoon & \astrosun\astrosun\fullmoon & \astrosun\astrosun\leftmoon & \astrosun\astrosun\fullmoon\\
0.5 &-16& \fullmoon\astrosun\leftmoon & \fullmoon\fullmoon\fullmoon & \fullmoon\fullmoon\fullmoon & \astrosun\astrosun\leftmoon & \fullmoon\fullmoon\leftmoon & \fullmoon\astrosun\fullmoon & \astrosun\astrosun\leftmoon & \fullmoon\fullmoon\leftmoon & \fullmoon\newmoon\leftmoon\\
\hline
\enddata
\tablecomments{ Upper nine blocks: $\mathcal{A}_{{\rm min}} > 1\ ($open, $ \fullmoon),\ 1 \geq \mathcal{A}_{{\rm min}} \geq 0.5\ ($dotted, $ \astrosun),\ \mathcal{A}_{{\rm min}} < 0.5\ ($closed, $ \newmoon)$; lower nine blocks: log$(\mathcal{T}) < 5\ ($crescent, $ \leftmoon)$, $5 <$ log$(\mathcal{T}) < 6\ ($open, $ \fullmoon)$, $6 <$ log$(\mathcal{T}) < 7\ ($dotted, $ \astrosun)$, log$(\mathcal{T}) > 7\ ($closed, $ \newmoon)$}
\end{deluxetable*}

%% file: moon_tmc1_low_er_rev.tex
\begin{deluxetable*}{cc|ccc|ccc|ccc}
\tabletypesize{\tiny}
\tablecaption{$\mathcal{A}_{{\rm min}}(\mathcal{T})$ and log($\mathcal{T}$) for Groups G, C, and S; Model B \label{ab}}
\tablehead{ & $a$ = & 0.01 & 0.01 & 0.01 & 0.03 & 0.03 & 0.03 & 0.10 & 0.10 & 0.10 \\
$db$ & log($\zeta$) $\backslash$ log($n$) = & 5 & 4.5 & 4 & 5 & 4.5 & 4 & 5 & 4.5 & 4
}
\decimals
\startdata
0.3 & -17 & \fullmoon\fullmoon\fullmoon & \fullmoon\fullmoon\astrosun & \fullmoon\astrosun\newmoon & \fullmoon\fullmoon\astrosun & \fullmoon\fullmoon\newmoon & \astrosun\newmoon\newmoon & \fullmoon\fullmoon\newmoon & \astrosun\astrosun\newmoon & \astrosun\newmoon\newmoon\\
0.3 &-16.5 & \fullmoon\fullmoon\astrosun & \fullmoon\astrosun\newmoon & \fullmoon\newmoon\newmoon & \fullmoon\fullmoon\newmoon & \astrosun\astrosun\newmoon & \astrosun\newmoon\newmoon & \astrosun\astrosun\newmoon & \astrosun\newmoon\newmoon & \astrosun\newmoon\newmoon\\
0.3 & -16 & \fullmoon\astrosun\newmoon & \fullmoon\newmoon\newmoon & \fullmoon\newmoon\astrosun & \astrosun\astrosun\newmoon & \astrosun\newmoon\newmoon & \astrosun\newmoon\newmoon & \astrosun\newmoon\newmoon & \astrosun\newmoon\newmoon & \astrosun\newmoon\newmoon\\
\hline
0.4 & 17 & \fullmoon\fullmoon\fullmoon & \fullmoon\fullmoon\astrosun & \fullmoon\astrosun\newmoon & \fullmoon\fullmoon\astrosun & \fullmoon\fullmoon\newmoon & \astrosun\newmoon\newmoon & \fullmoon\fullmoon\newmoon & \astrosun\astrosun\newmoon & \astrosun\newmoon\newmoon\\
0.4 & -16.5 & \fullmoon\fullmoon\astrosun & \fullmoon\astrosun\newmoon & \fullmoon\newmoon\newmoon & \fullmoon\fullmoon\newmoon & \astrosun\newmoon\newmoon & \astrosun\newmoon\newmoon & \astrosun\astrosun\newmoon & \astrosun\newmoon\newmoon & \astrosun\newmoon\newmoon\\
0.4 & -16 & \fullmoon\astrosun\newmoon & \fullmoon\newmoon\newmoon & \fullmoon\newmoon\astrosun & \astrosun\astrosun\newmoon & \astrosun\newmoon\newmoon & \astrosun\newmoon\newmoon & \astrosun\newmoon\newmoon & \astrosun\newmoon\newmoon & \astrosun\newmoon\newmoon\\
\hline
0.5 &-17 & \fullmoon\fullmoon\fullmoon & \fullmoon\fullmoon\astrosun & \fullmoon\astrosun\newmoon & \fullmoon\fullmoon\astrosun & \fullmoon\fullmoon\newmoon & \astrosun\newmoon\newmoon & \fullmoon\fullmoon\newmoon & \astrosun\astrosun\newmoon & \astrosun\newmoon\newmoon\\
0.5 & -16.5& \fullmoon\fullmoon\astrosun & \fullmoon\astrosun\newmoon & \fullmoon\newmoon\newmoon & \fullmoon\fullmoon\newmoon & \astrosun\newmoon\newmoon & \astrosun\newmoon\newmoon & \astrosun\astrosun\newmoon & \astrosun\newmoon\newmoon & \astrosun\newmoon\newmoon\\
0.5 &-16& \fullmoon\astrosun\newmoon & \fullmoon\newmoon\newmoon & \fullmoon\newmoon\astrosun & \astrosun\astrosun\newmoon & \astrosun\newmoon\newmoon & \astrosun\newmoon\newmoon & \astrosun\newmoon\newmoon & \astrosun\newmoon\newmoon & \astrosun\newmoon\newmoon\\
\hline
\hline
0.3 & -17 & \leftmoon\leftmoon\fullmoon & \fullmoon\newmoon\fullmoon & \newmoon\newmoon\fullmoon & \leftmoon\leftmoon\fullmoon & \newmoon\newmoon\fullmoon & \newmoon\newmoon\fullmoon & \newmoon\newmoon\fullmoon & \newmoon\newmoon\fullmoon & \astrosun\newmoon\fullmoon\\
0.3 &-16.5 & \leftmoon\astrosun\fullmoon & \astrosun\astrosun\fullmoon & \astrosun\astrosun\fullmoon & \astrosun\astrosun\fullmoon & \astrosun\astrosun\fullmoon & \astrosun\astrosun\fullmoon & \astrosun\astrosun\fullmoon & \astrosun\astrosun\leftmoon & \astrosun\newmoon\fullmoon\\
0.3 & -16 & \astrosun\astrosun\leftmoon & \fullmoon\fullmoon\fullmoon & \fullmoon\fullmoon\fullmoon & \astrosun\astrosun\leftmoon & \fullmoon\fullmoon\leftmoon & \fullmoon\astrosun\fullmoon & \astrosun\astrosun\leftmoon & \fullmoon\astrosun\leftmoon & \fullmoon\newmoon\leftmoon\\
\hline
0.4 & 17 & \leftmoon\leftmoon\fullmoon & \fullmoon\newmoon\fullmoon & \newmoon\newmoon\fullmoon & \leftmoon\leftmoon\fullmoon & \newmoon\newmoon\fullmoon & \newmoon\newmoon\fullmoon & \newmoon\newmoon\fullmoon & \newmoon\newmoon\fullmoon & \astrosun\newmoon\fullmoon\\
0.4 & -16.5 & \leftmoon\astrosun\fullmoon & \astrosun\astrosun\fullmoon & \astrosun\astrosun\fullmoon & \astrosun\astrosun\fullmoon & \astrosun\astrosun\fullmoon & \astrosun\astrosun\fullmoon & \astrosun\astrosun\fullmoon & \astrosun\astrosun\leftmoon & \astrosun\newmoon\fullmoon\\
0.4 & -16 & \astrosun\astrosun\leftmoon & \fullmoon\fullmoon\fullmoon & \fullmoon\fullmoon\fullmoon & \astrosun\astrosun\leftmoon & \fullmoon\fullmoon\leftmoon & \fullmoon\astrosun\fullmoon & \astrosun\astrosun\leftmoon & \fullmoon\astrosun\leftmoon & \fullmoon\newmoon\leftmoon\\
\hline
0.5 &-17 & \leftmoon\leftmoon\fullmoon & \fullmoon\newmoon\fullmoon & \newmoon\newmoon\fullmoon & \leftmoon\leftmoon\fullmoon & \newmoon\newmoon\fullmoon & \newmoon\newmoon\fullmoon & \newmoon\newmoon\fullmoon & \newmoon\newmoon\fullmoon & \astrosun\newmoon\fullmoon\\
0.5 & -16.5& \leftmoon\astrosun\fullmoon & \astrosun\astrosun\fullmoon & \astrosun\astrosun\fullmoon & \astrosun\astrosun\fullmoon & \astrosun\astrosun\fullmoon & \astrosun\astrosun\fullmoon & \astrosun\astrosun\fullmoon & \astrosun\astrosun\leftmoon & \astrosun\newmoon\fullmoon\\
0.5 &-16& \astrosun\astrosun\leftmoon & \fullmoon\fullmoon\fullmoon & \fullmoon\fullmoon\fullmoon & \astrosun\astrosun\leftmoon & \fullmoon\fullmoon\leftmoon & \fullmoon\astrosun\fullmoon & \astrosun\astrosun\leftmoon & \fullmoon\astrosun\leftmoon & \fullmoon\newmoon\leftmoon\\
\hline
\enddata
\tablecomments{ Upper nine blocks: $\mathcal{A}_{{\rm min}} > 1\ ($open, $ \fullmoon),\ 1 \geq \mathcal{A}_{{\rm min}} \geq 0.5\ ($dotted, $ \astrosun),\ \mathcal{A}_{{\rm min}} < 0.5\ ($closed, $ \newmoon)$; lower nine blocks: log$(\mathcal{T}) < 5\ ($crescent, $ \leftmoon)$, $5 <$ log$(\mathcal{T}) < 6\ ($open, $ \fullmoon)$, $6 <$ log$(\mathcal{T}) < 7\ ($dotted, $ \astrosun)$, log$(\mathcal{T}) > 7\ ($closed, $ \newmoon)$}
\end{deluxetable*}

%% file: moon_tmc1_high_rev.tex
\begin{deluxetable*}{cc|ccc|ccc|ccc}
\tabletypesize{\tiny}
\tablecaption{$\mathcal{A}_{{\rm min}}(\mathcal{T})$ and log($\mathcal{T}$) for Groups G, C, and S; Model C \label{ac}}
\tablehead{ & $a$ = & 0.01 & 0.01 & 0.01 & 0.03 & 0.03 & 0.03 & 0.10 & 0.10 & 0.10 \\
$db$ & log($\zeta$) $\backslash$ log($n$) = & 5 & 4.5 & 4 & 5 & 4.5 & 4 & 5 & 4.5 & 4
}
\decimals
\startdata
0.3 & -17 & \fullmoon\newmoon\fullmoon & \astrosun\newmoon\fullmoon & \astrosun\newmoon\astrosun & \astrosun\newmoon\fullmoon & \astrosun\newmoon\astrosun & \astrosun\newmoon\newmoon & \astrosun\newmoon\astrosun & \astrosun\newmoon\newmoon & \astrosun\newmoon\newmoon\\
0.3 &-16.5 & \astrosun\newmoon\fullmoon & \astrosun\newmoon\astrosun & \astrosun\newmoon\astrosun & \astrosun\newmoon\astrosun & \astrosun\newmoon\newmoon & \astrosun\newmoon\newmoon & \astrosun\newmoon\newmoon & \astrosun\newmoon\newmoon & \astrosun\newmoon\newmoon\\
0.3 & -16 & \astrosun\newmoon\astrosun & \astrosun\newmoon\astrosun & \fullmoon\newmoon\astrosun & \astrosun\newmoon\newmoon & \astrosun\newmoon\newmoon & \fullmoon\newmoon\astrosun & \astrosun\newmoon\newmoon & \astrosun\newmoon\newmoon & \astrosun\newmoon\newmoon\\
\hline
0.4 & 17 & \fullmoon\newmoon\fullmoon & \astrosun\newmoon\fullmoon & \astrosun\newmoon\astrosun & \astrosun\newmoon\fullmoon & \astrosun\newmoon\astrosun & \astrosun\newmoon\newmoon & \astrosun\newmoon\astrosun & \astrosun\newmoon\newmoon & \astrosun\newmoon\newmoon\\
0.4 & -16.5 & \astrosun\newmoon\fullmoon & \astrosun\newmoon\astrosun & \fullmoon\newmoon\astrosun & \astrosun\newmoon\astrosun & \astrosun\newmoon\newmoon & \astrosun\newmoon\newmoon & \astrosun\newmoon\newmoon & \astrosun\newmoon\newmoon & \astrosun\newmoon\newmoon\\
0.4 & -16 & \astrosun\newmoon\astrosun & \fullmoon\newmoon\astrosun & \fullmoon\newmoon\astrosun & \astrosun\newmoon\newmoon & \astrosun\newmoon\newmoon & \fullmoon\newmoon\astrosun & \astrosun\newmoon\newmoon & \astrosun\newmoon\newmoon & \astrosun\newmoon\newmoon\\
\hline
0.5 &-17 & \fullmoon\newmoon\fullmoon & \astrosun\newmoon\fullmoon & \astrosun\newmoon\astrosun & \astrosun\newmoon\fullmoon & \astrosun\newmoon\astrosun & \astrosun\newmoon\newmoon & \astrosun\newmoon\astrosun & \astrosun\newmoon\newmoon & \astrosun\newmoon\newmoon\\
0.5 & -16.5& \astrosun\newmoon\fullmoon & \astrosun\newmoon\astrosun & \fullmoon\newmoon\astrosun & \astrosun\newmoon\astrosun & \astrosun\newmoon\newmoon & \astrosun\newmoon\newmoon & \astrosun\newmoon\newmoon & \astrosun\newmoon\newmoon & \astrosun\newmoon\newmoon\\
0.5 &-16& \astrosun\newmoon\fullmoon & \fullmoon\newmoon\astrosun & \fullmoon\newmoon\astrosun & \astrosun\newmoon\newmoon & \astrosun\newmoon\newmoon & \fullmoon\newmoon\astrosun & \astrosun\newmoon\newmoon & \astrosun\newmoon\newmoon & \astrosun\newmoon\newmoon\\
\hline
\hline
0.3 & -17 & \fullmoon\leftmoon\fullmoon & \fullmoon\fullmoon\fullmoon & \astrosun\fullmoon\fullmoon & \fullmoon\leftmoon\fullmoon & \fullmoon\fullmoon\fullmoon & \astrosun\fullmoon\fullmoon & \fullmoon\leftmoon\fullmoon & \fullmoon\fullmoon\fullmoon & \astrosun\fullmoon\fullmoon\\
0.3 &-16.5 & \fullmoon\leftmoon\fullmoon & \fullmoon\fullmoon\fullmoon & \fullmoon\fullmoon\fullmoon & \fullmoon\leftmoon\fullmoon & \fullmoon\fullmoon\fullmoon & \astrosun\fullmoon\fullmoon & \fullmoon\leftmoon\fullmoon & \fullmoon\leftmoon\fullmoon & \newmoon\fullmoon\fullmoon\\
0.3 & -16 & \fullmoon\leftmoon\leftmoon & \fullmoon\leftmoon\leftmoon & \fullmoon\leftmoon\fullmoon & \fullmoon\leftmoon\leftmoon & \fullmoon\leftmoon\leftmoon & \astrosun\leftmoon\leftmoon & \fullmoon\leftmoon\leftmoon & \astrosun\leftmoon\leftmoon & \newmoon\leftmoon\leftmoon\\
\hline
0.4 & 17 & \fullmoon\leftmoon\fullmoon & \fullmoon\fullmoon\fullmoon & \astrosun\fullmoon\fullmoon & \fullmoon\leftmoon\fullmoon & \fullmoon\fullmoon\fullmoon & \astrosun\fullmoon\fullmoon & \fullmoon\leftmoon\fullmoon & \fullmoon\fullmoon\fullmoon & \astrosun\fullmoon\fullmoon\\
0.4 & -16.5 & \fullmoon\leftmoon\fullmoon & \fullmoon\fullmoon\fullmoon & \fullmoon\fullmoon\fullmoon & \fullmoon\leftmoon\fullmoon & \fullmoon\fullmoon\fullmoon & \astrosun\fullmoon\fullmoon & \fullmoon\leftmoon\fullmoon & \fullmoon\leftmoon\fullmoon & \newmoon\fullmoon\fullmoon\\
0.4 & -16 & \fullmoon\leftmoon\leftmoon & \fullmoon\leftmoon\leftmoon & \fullmoon\leftmoon\fullmoon & \fullmoon\leftmoon\leftmoon & \fullmoon\leftmoon\leftmoon & \astrosun\leftmoon\leftmoon & \fullmoon\leftmoon\leftmoon & \astrosun\leftmoon\leftmoon & \newmoon\leftmoon\leftmoon\\
\hline
0.5 &-17 & \fullmoon\leftmoon\newmoon & \fullmoon\fullmoon\fullmoon & \astrosun\fullmoon\fullmoon & \fullmoon\leftmoon\newmoon & \fullmoon\fullmoon\fullmoon & \astrosun\fullmoon\fullmoon & \fullmoon\leftmoon\fullmoon & \fullmoon\fullmoon\fullmoon & \astrosun\fullmoon\fullmoon\\
0.5 & -16.5& \fullmoon\leftmoon\fullmoon & \fullmoon\fullmoon\fullmoon & \fullmoon\fullmoon\fullmoon & \fullmoon\leftmoon\fullmoon & \fullmoon\fullmoon\fullmoon & \astrosun\fullmoon\fullmoon & \fullmoon\leftmoon\fullmoon & \fullmoon\leftmoon\fullmoon & \newmoon\fullmoon\fullmoon\\
0.5 &-16& \fullmoon\leftmoon\leftmoon & \fullmoon\leftmoon\leftmoon & \astrosun\leftmoon\fullmoon & \fullmoon\leftmoon\leftmoon & \fullmoon\leftmoon\leftmoon & \astrosun\leftmoon\leftmoon & \fullmoon\leftmoon\leftmoon & \astrosun\leftmoon\leftmoon & \newmoon\leftmoon\leftmoon\\
\hline
\enddata
\tablecomments{ Upper nine blocks: $\mathcal{A}_{{\rm min}} > 1\ ($open, $ \fullmoon),\ 1 \geq \mathcal{A}_{{\rm min}} \geq 0.5\ ($dotted, $ \astrosun),\ \mathcal{A}_{{\rm min}} < 0.5\ ($closed, $ \newmoon)$; lower nine blocks: log$(\mathcal{T}) < 5\ ($crescent, $ \leftmoon)$, $5 <$ log$(\mathcal{T}) < 6\ ($open, $ \fullmoon)$, $6 <$ log$(\mathcal{T}) < 7\ ($dotted, $ \astrosun)$, log$(\mathcal{T}) > 7\ ($closed, $ \newmoon)$}
\end{deluxetable*}

%% file: moon_tmc1_high_er_rev.tex
\begin{deluxetable*}{cc|ccc|ccc|ccc}
\tabletypesize{\tiny}
\tablecaption{$\mathcal{A}_{{\rm min}}(\mathcal{T})$ and log($\mathcal{T}$) for Groups G, C, and S; Model D \label{ad}}
\tablehead{ & $a$ = & 0.01 & 0.01 & 0.01 & 0.03 & 0.03 & 0.03 & 0.10 & 0.10 & 0.10 \\
$db$ & log($\zeta$) $\backslash$ log($n$) = & 5 & 4.5 & 4 & 5 & 4.5 & 4 & 5 & 4.5 & 4
}
\decimals
\startdata
0.3 & -17 & \astrosun\newmoon\astrosun & \astrosun\newmoon\astrosun & \astrosun\newmoon\astrosun & \astrosun\newmoon\newmoon & \astrosun\newmoon\newmoon & \astrosun\newmoon\newmoon & \astrosun\newmoon\newmoon & \astrosun\newmoon\newmoon & \astrosun\newmoon\newmoon\\
0.3 &-16.5 & \astrosun\newmoon\astrosun & \astrosun\newmoon\astrosun & \astrosun\newmoon\astrosun & \astrosun\newmoon\newmoon & \astrosun\newmoon\newmoon & \astrosun\newmoon\newmoon & \astrosun\newmoon\newmoon & \astrosun\newmoon\newmoon & \astrosun\newmoon\newmoon\\
0.3 & -16 & \astrosun\newmoon\astrosun & \astrosun\newmoon\astrosun & \fullmoon\newmoon\astrosun & \astrosun\newmoon\newmoon & \astrosun\newmoon\newmoon & \fullmoon\newmoon\newmoon & \astrosun\newmoon\newmoon & \astrosun\newmoon\newmoon & \astrosun\newmoon\newmoon\\
\hline
0.4 & 17 & \astrosun\astrosun\fullmoon & \astrosun\newmoon\astrosun & \astrosun\newmoon\astrosun & \astrosun\newmoon\astrosun & \astrosun\newmoon\newmoon & \astrosun\newmoon\newmoon & \astrosun\newmoon\newmoon & \astrosun\newmoon\newmoon & \astrosun\newmoon\newmoon\\
0.4 & -16.5 & \astrosun\newmoon\astrosun & \astrosun\newmoon\astrosun & \astrosun\newmoon\astrosun & \astrosun\newmoon\newmoon & \astrosun\newmoon\newmoon & \astrosun\newmoon\newmoon & \astrosun\newmoon\newmoon & \astrosun\newmoon\newmoon & \astrosun\newmoon\newmoon\\
0.4 & -16 & \astrosun\newmoon\astrosun & \astrosun\newmoon\astrosun & \fullmoon\newmoon\astrosun & \astrosun\newmoon\newmoon & \astrosun\newmoon\newmoon & \fullmoon\newmoon\newmoon & \astrosun\newmoon\newmoon & \astrosun\newmoon\newmoon & \astrosun\newmoon\newmoon\\
\hline
0.5 &-17 & \astrosun\astrosun\fullmoon & \astrosun\newmoon\astrosun & \astrosun\newmoon\astrosun & \astrosun\newmoon\astrosun & \astrosun\newmoon\newmoon & \astrosun\newmoon\newmoon & \astrosun\newmoon\newmoon & \astrosun\newmoon\newmoon & \astrosun\newmoon\newmoon\\
0.5 & -16.5& \astrosun\newmoon\astrosun & \astrosun\newmoon\astrosun & \astrosun\newmoon\astrosun & \astrosun\newmoon\newmoon & \astrosun\newmoon\newmoon & \astrosun\newmoon\newmoon & \astrosun\newmoon\newmoon & \astrosun\newmoon\newmoon & \astrosun\newmoon\newmoon\\
0.5 &-16& \astrosun\newmoon\astrosun & \astrosun\newmoon\astrosun & \fullmoon\newmoon\astrosun & \astrosun\newmoon\newmoon & \astrosun\newmoon\newmoon & \fullmoon\newmoon\newmoon & \astrosun\newmoon\newmoon & \astrosun\newmoon\newmoon & \astrosun\newmoon\newmoon\\
\hline
\hline
0.3 & -17 & \fullmoon\fullmoon\fullmoon & \fullmoon\fullmoon\fullmoon & \fullmoon\fullmoon\fullmoon & \fullmoon\fullmoon\fullmoon & \fullmoon\fullmoon\fullmoon & \astrosun\fullmoon\fullmoon & \fullmoon\fullmoon\fullmoon & \fullmoon\fullmoon\fullmoon & \astrosun\astrosun\fullmoon\\
0.3 &-16.5 & \fullmoon\leftmoon\fullmoon & \fullmoon\fullmoon\fullmoon & \fullmoon\fullmoon\fullmoon & \fullmoon\leftmoon\fullmoon & \fullmoon\fullmoon\fullmoon & \fullmoon\fullmoon\fullmoon & \fullmoon\leftmoon\fullmoon & \fullmoon\fullmoon\leftmoon & \newmoon\fullmoon\fullmoon\\
0.3 & -16 & \fullmoon\leftmoon\leftmoon & \fullmoon\leftmoon\leftmoon & \fullmoon\leftmoon\fullmoon & \fullmoon\leftmoon\leftmoon & \fullmoon\leftmoon\leftmoon & \astrosun\leftmoon\leftmoon & \fullmoon\fullmoon\leftmoon & \astrosun\leftmoon\leftmoon & \newmoon\leftmoon\leftmoon\\
\hline
0.4 & 17 & \fullmoon\fullmoon\fullmoon & \fullmoon\fullmoon\fullmoon & \fullmoon\fullmoon\fullmoon & \fullmoon\fullmoon\fullmoon & \fullmoon\fullmoon\fullmoon & \astrosun\fullmoon\fullmoon & \fullmoon\fullmoon\fullmoon & \fullmoon\fullmoon\fullmoon & \astrosun\astrosun\fullmoon\\
0.4 & -16.5 & \fullmoon\leftmoon\fullmoon & \fullmoon\fullmoon\fullmoon & \fullmoon\fullmoon\fullmoon & \fullmoon\leftmoon\fullmoon & \fullmoon\fullmoon\fullmoon & \fullmoon\fullmoon\fullmoon & \fullmoon\leftmoon\fullmoon & \fullmoon\fullmoon\leftmoon & \newmoon\fullmoon\fullmoon\\
0.4 & -16 & \fullmoon\leftmoon\leftmoon & \fullmoon\leftmoon\leftmoon & \fullmoon\leftmoon\fullmoon & \fullmoon\leftmoon\leftmoon & \fullmoon\leftmoon\leftmoon & \astrosun\leftmoon\leftmoon & \fullmoon\fullmoon\leftmoon & \astrosun\leftmoon\leftmoon & \newmoon\leftmoon\leftmoon\\
\hline
0.5 &-17 & \fullmoon\fullmoon\fullmoon & \fullmoon\fullmoon\fullmoon & \fullmoon\fullmoon\fullmoon & \fullmoon\fullmoon\fullmoon & \fullmoon\fullmoon\fullmoon & \astrosun\fullmoon\fullmoon & \fullmoon\fullmoon\fullmoon & \fullmoon\fullmoon\fullmoon & \astrosun\astrosun\fullmoon\\
0.5 & -16.5& \fullmoon\leftmoon\fullmoon & \fullmoon\fullmoon\fullmoon & \fullmoon\fullmoon\fullmoon & \fullmoon\leftmoon\fullmoon & \fullmoon\fullmoon\fullmoon & \fullmoon\fullmoon\fullmoon & \fullmoon\leftmoon\fullmoon & \fullmoon\fullmoon\leftmoon & \newmoon\fullmoon\fullmoon\\
0.5 &-16& \fullmoon\leftmoon\leftmoon & \fullmoon\leftmoon\leftmoon & \fullmoon\leftmoon\fullmoon & \fullmoon\leftmoon\leftmoon & \fullmoon\leftmoon\leftmoon & \astrosun\leftmoon\leftmoon & \fullmoon\fullmoon\leftmoon & \astrosun\leftmoon\leftmoon & \newmoon\leftmoon\leftmoon\\
\hline
\enddata
\tablecomments{ Upper nine blocks: $\mathcal{A}_{{\rm min}} > 1\ ($open, $ \fullmoon),\ 1 \geq \mathcal{A}_{{\rm min}} \geq 0.5\ ($dotted, $ \astrosun),\ \mathcal{A}_{{\rm min}} < 0.5\ ($closed, $ \newmoon)$; lower nine blocks: log$(\mathcal{T}) < 5\ ($crescent, $ \leftmoon)$, $5 <$ log$(\mathcal{T}) < 6\ ($open, $ \fullmoon)$, $6 <$ log$(\mathcal{T}) < 7\ ($dotted, $ \astrosun)$, log$(\mathcal{T}) > 7\ ($closed, $ \newmoon)$}
\end{deluxetable*}

%% file: moon_sigma_tmc1_low_rev.tex
\begin{deluxetable*}{cc|ccc|ccc|ccc}
\tabletypesize{\tiny}
\tablecaption{$\sigma_{{\rm min}}(\mathcal{T})$ and log($\mathcal{T}$) for Groups G, C, and S; Model A \label{sa}}
\tablehead{ & $a$ = & 0.01 & 0.01 & 0.01 & 0.03 & 0.03 & 0.03 & 0.10 & 0.10 & 0.10 \\
$db$ & log($\zeta$) $\backslash$ log($n$) = & 5 & 4.5 & 4 & 5 & 4.5 & 4 & 5 & 4.5 & 4
}
\decimals
\startdata
0.3 & -17 & \fullmoon\fullmoon\fullmoon & \fullmoon\fullmoon\astrosun & \fullmoon\fullmoon\newmoon & \fullmoon\fullmoon\astrosun & \fullmoon\fullmoon\newmoon & \fullmoon\fullmoon\newmoon & \fullmoon\fullmoon\newmoon & \fullmoon\fullmoon\newmoon & \fullmoon\fullmoon\newmoon\\
0.3 &-16.5 & \fullmoon\fullmoon\astrosun & \fullmoon\fullmoon\newmoon & \fullmoon\fullmoon\newmoon & \fullmoon\fullmoon\newmoon & \fullmoon\fullmoon\newmoon & \fullmoon\fullmoon\newmoon & \fullmoon\fullmoon\newmoon & \fullmoon\fullmoon\newmoon & \fullmoon\fullmoon\newmoon\\
0.3 & -16 & \fullmoon\fullmoon\newmoon & \fullmoon\fullmoon\newmoon & \fullmoon\fullmoon\newmoon & \fullmoon\fullmoon\newmoon & \fullmoon\fullmoon\newmoon & \fullmoon\fullmoon\newmoon & \fullmoon\fullmoon\newmoon & \fullmoon\fullmoon\newmoon & \fullmoon\fullmoon\newmoon\\
\hline
0.4 & 17 & \fullmoon\fullmoon\fullmoon & \fullmoon\fullmoon\astrosun & \fullmoon\fullmoon\newmoon & \fullmoon\fullmoon\astrosun & \fullmoon\fullmoon\newmoon & \fullmoon\fullmoon\newmoon & \fullmoon\fullmoon\newmoon & \fullmoon\fullmoon\newmoon & \fullmoon\fullmoon\newmoon\\
0.4 & -16.5 & \fullmoon\fullmoon\astrosun & \fullmoon\fullmoon\newmoon & \fullmoon\fullmoon\newmoon & \fullmoon\fullmoon\newmoon & \fullmoon\fullmoon\newmoon & \fullmoon\fullmoon\newmoon & \fullmoon\fullmoon\newmoon & \fullmoon\fullmoon\newmoon & \fullmoon\fullmoon\newmoon\\
0.4 & -16 & \fullmoon\fullmoon\newmoon & \fullmoon\fullmoon\newmoon & \fullmoon\fullmoon\newmoon & \fullmoon\fullmoon\newmoon & \fullmoon\fullmoon\newmoon & \fullmoon\fullmoon\newmoon & \fullmoon\fullmoon\newmoon & \fullmoon\fullmoon\newmoon & \fullmoon\fullmoon\newmoon\\
\hline
0.5 &-17 & \fullmoon\fullmoon\fullmoon & \fullmoon\fullmoon\astrosun & \fullmoon\fullmoon\newmoon & \fullmoon\fullmoon\astrosun & \fullmoon\fullmoon\newmoon & \fullmoon\fullmoon\newmoon & \fullmoon\fullmoon\newmoon & \fullmoon\fullmoon\newmoon & \fullmoon\fullmoon\newmoon\\
0.5 & -16.5& \fullmoon\fullmoon\astrosun & \fullmoon\fullmoon\newmoon & \fullmoon\fullmoon\newmoon & \fullmoon\fullmoon\newmoon & \fullmoon\fullmoon\newmoon & \fullmoon\fullmoon\newmoon & \fullmoon\fullmoon\newmoon & \fullmoon\fullmoon\newmoon & \fullmoon\fullmoon\newmoon\\
0.5 &-16& \fullmoon\fullmoon\newmoon & \fullmoon\fullmoon\newmoon & \fullmoon\fullmoon\newmoon & \fullmoon\fullmoon\newmoon & \fullmoon\fullmoon\newmoon & \fullmoon\fullmoon\newmoon & \fullmoon\fullmoon\newmoon & \fullmoon\fullmoon\newmoon & \fullmoon\fullmoon\newmoon\\
\hline
\hline
0.3 & -17 & \leftmoon\leftmoon\fullmoon & \newmoon\newmoon\fullmoon & \newmoon\newmoon\fullmoon & \leftmoon\leftmoon\fullmoon & \newmoon\newmoon\fullmoon & \newmoon\newmoon\astrosun & \newmoon\leftmoon\fullmoon & \newmoon\newmoon\fullmoon & \astrosun\astrosun\fullmoon\\
0.3 &-16.5 & \astrosun\astrosun\fullmoon & \astrosun\astrosun\fullmoon & \astrosun\astrosun\fullmoon & \astrosun\astrosun\fullmoon & \astrosun\astrosun\fullmoon & \astrosun\astrosun\fullmoon & \astrosun\astrosun\leftmoon & \astrosun\astrosun\leftmoon & \astrosun\newmoon\fullmoon\\
0.3 & -16 & \astrosun\astrosun\leftmoon & \fullmoon\fullmoon\fullmoon & \fullmoon\astrosun\fullmoon & \astrosun\astrosun\leftmoon & \fullmoon\fullmoon\leftmoon & \astrosun\astrosun\fullmoon & \fullmoon\astrosun\leftmoon & \fullmoon\astrosun\leftmoon & \newmoon\newmoon\leftmoon\\
\hline
0.4 & 17 & \leftmoon\leftmoon\fullmoon & \fullmoon\fullmoon\fullmoon & \newmoon\newmoon\fullmoon & \leftmoon\leftmoon\fullmoon & \newmoon\newmoon\fullmoon & \newmoon\newmoon\astrosun & \newmoon\leftmoon\fullmoon & \newmoon\newmoon\fullmoon & \astrosun\astrosun\fullmoon\\
0.4 & -16.5 & \fullmoon\fullmoon\fullmoon & \astrosun\astrosun\fullmoon & \astrosun\astrosun\fullmoon & \astrosun\astrosun\fullmoon & \astrosun\astrosun\fullmoon & \astrosun\astrosun\fullmoon & \astrosun\astrosun\fullmoon & \astrosun\astrosun\leftmoon & \astrosun\astrosun\fullmoon\\
0.4 & -16 & \astrosun\astrosun\leftmoon & \fullmoon\fullmoon\fullmoon & \fullmoon\fullmoon\fullmoon & \astrosun\astrosun\leftmoon & \fullmoon\fullmoon\leftmoon & \fullmoon\astrosun\fullmoon & \fullmoon\astrosun\leftmoon & \fullmoon\fullmoon\leftmoon & \fullmoon\newmoon\leftmoon\\
\hline
0.5 &-17 & \leftmoon\leftmoon\fullmoon & \fullmoon\fullmoon\fullmoon & \astrosun\newmoon\fullmoon & \leftmoon\leftmoon\newmoon & \newmoon\newmoon\fullmoon & \newmoon\newmoon\astrosun & \newmoon\leftmoon\fullmoon & \newmoon\newmoon\fullmoon & \astrosun\astrosun\fullmoon\\
0.5 & -16.5& \fullmoon\fullmoon\fullmoon & \fullmoon\astrosun\fullmoon & \astrosun\astrosun\fullmoon & \astrosun\astrosun\fullmoon & \astrosun\astrosun\fullmoon & \astrosun\astrosun\fullmoon & \astrosun\astrosun\leftmoon & \astrosun\astrosun\leftmoon & \astrosun\astrosun\fullmoon\\
0.5 &-16& \fullmoon\astrosun\leftmoon & \fullmoon\fullmoon\fullmoon & \fullmoon\fullmoon\fullmoon & \astrosun\astrosun\leftmoon & \fullmoon\fullmoon\leftmoon & \fullmoon\astrosun\fullmoon & \fullmoon\astrosun\leftmoon & \fullmoon\fullmoon\leftmoon & \fullmoon\newmoon\leftmoon\\
\hline
\enddata
\tablecomments{ Upper nine blocks: $\sigma_{{\rm min}} > 1\ ($open, $ \fullmoon),\ 1 \geq \sigma_{{\rm min}} \geq 0.5\ ($dotted, $ \astrosun),\ \sigma_{{\rm min}} < 0.5\ ($closed, $ \newmoon)$; lower nine blocks: log$(\mathcal{T}) < 5\ ($crescent, $ \leftmoon)$, $5 <$ log$(\mathcal{T}) < 6\ ($open, $ \fullmoon)$, $6 <$ log$(\mathcal{T}) < 7\ ($dotted, $ \astrosun)$, log$(\mathcal{T}) > 7\ ($closed, $ \newmoon)$}
\end{deluxetable*}

%% file: moon_sigma_tmc1_low_er_rev.tex
\begin{deluxetable*}{cc|ccc|ccc|ccc}
\tabletypesize{\tiny}
\tablecaption{$\sigma_{{\rm min}}(\mathcal{T})$ and log($\mathcal{T}$) for Groups G, C, and S; Model B \label{sb}}
\tablehead{ & $a$ = & 0.01 & 0.01 & 0.01 & 0.03 & 0.03 & 0.03 & 0.10 & 0.10 & 0.10 \\
$db$ & log($\zeta$) $\backslash$ log($n$) = & 5 & 4.5 & 4 & 5 & 4.5 & 4 & 5 & 4.5 & 4
}
\decimals
\startdata
0.3 & -17 & \fullmoon\fullmoon\astrosun & \fullmoon\fullmoon\newmoon & \fullmoon\fullmoon\newmoon & \fullmoon\fullmoon\newmoon & \fullmoon\fullmoon\newmoon & \fullmoon\fullmoon\newmoon & \fullmoon\fullmoon\newmoon & \fullmoon\fullmoon\newmoon & \fullmoon\fullmoon\newmoon\\
0.3 &-16.5 & \fullmoon\fullmoon\newmoon & \fullmoon\fullmoon\newmoon & \fullmoon\fullmoon\newmoon & \fullmoon\fullmoon\newmoon & \fullmoon\fullmoon\newmoon & \fullmoon\fullmoon\newmoon & \fullmoon\fullmoon\newmoon & \fullmoon\fullmoon\newmoon & \fullmoon\fullmoon\newmoon\\
0.3 & -16 & \fullmoon\fullmoon\newmoon & \fullmoon\fullmoon\newmoon & \fullmoon\fullmoon\newmoon & \fullmoon\fullmoon\newmoon & \fullmoon\fullmoon\newmoon & \fullmoon\fullmoon\newmoon & \fullmoon\fullmoon\newmoon & \fullmoon\fullmoon\newmoon & \fullmoon\fullmoon\newmoon\\
\hline
0.4 & 17 & \fullmoon\fullmoon\astrosun & \fullmoon\fullmoon\newmoon & \fullmoon\fullmoon\newmoon & \fullmoon\fullmoon\newmoon & \fullmoon\fullmoon\newmoon & \fullmoon\fullmoon\newmoon & \fullmoon\fullmoon\newmoon & \fullmoon\fullmoon\newmoon & \fullmoon\fullmoon\newmoon\\
0.4 & -16.5 & \fullmoon\fullmoon\newmoon & \fullmoon\fullmoon\newmoon & \fullmoon\fullmoon\newmoon & \fullmoon\fullmoon\newmoon & \fullmoon\fullmoon\newmoon & \fullmoon\fullmoon\newmoon & \fullmoon\fullmoon\newmoon & \fullmoon\fullmoon\newmoon & \fullmoon\fullmoon\newmoon\\
0.4 & -16 & \fullmoon\fullmoon\newmoon & \fullmoon\fullmoon\newmoon & \fullmoon\fullmoon\newmoon & \fullmoon\fullmoon\newmoon & \fullmoon\fullmoon\newmoon & \fullmoon\fullmoon\newmoon & \fullmoon\fullmoon\newmoon & \fullmoon\fullmoon\newmoon & \fullmoon\fullmoon\newmoon\\
\hline
0.5 &-17 & \fullmoon\fullmoon\astrosun & \fullmoon\fullmoon\newmoon & \fullmoon\fullmoon\newmoon & \fullmoon\fullmoon\newmoon & \fullmoon\fullmoon\newmoon & \fullmoon\fullmoon\newmoon & \fullmoon\fullmoon\newmoon & \fullmoon\fullmoon\newmoon & \fullmoon\fullmoon\newmoon\\
0.5 & -16.5& \fullmoon\fullmoon\newmoon & \fullmoon\fullmoon\newmoon & \fullmoon\fullmoon\newmoon & \fullmoon\fullmoon\newmoon & \fullmoon\fullmoon\newmoon & \fullmoon\fullmoon\newmoon & \fullmoon\fullmoon\newmoon & \fullmoon\fullmoon\newmoon & \fullmoon\fullmoon\newmoon\\
0.5 &-16& \fullmoon\fullmoon\newmoon & \fullmoon\fullmoon\newmoon & \fullmoon\fullmoon\newmoon & \fullmoon\fullmoon\newmoon & \fullmoon\fullmoon\newmoon & \fullmoon\fullmoon\newmoon & \fullmoon\fullmoon\newmoon & \fullmoon\fullmoon\newmoon & \fullmoon\fullmoon\newmoon\\
\hline
\hline
0.3 & -17 & \leftmoon\leftmoon\fullmoon & \newmoon\newmoon\fullmoon & \newmoon\newmoon\fullmoon & \newmoon\leftmoon\fullmoon & \newmoon\newmoon\fullmoon & \newmoon\newmoon\fullmoon & \newmoon\newmoon\fullmoon & \newmoon\newmoon\fullmoon & \astrosun\astrosun\fullmoon\\
0.3 &-16.5 & \astrosun\astrosun\fullmoon & \astrosun\astrosun\fullmoon & \astrosun\astrosun\fullmoon & \astrosun\astrosun\fullmoon & \astrosun\astrosun\fullmoon & \astrosun\astrosun\fullmoon & \astrosun\astrosun\leftmoon & \astrosun\astrosun\leftmoon & \astrosun\newmoon\fullmoon\\
0.3 & -16 & \astrosun\astrosun\leftmoon & \fullmoon\fullmoon\fullmoon & \fullmoon\astrosun\fullmoon & \astrosun\astrosun\leftmoon & \fullmoon\fullmoon\leftmoon & \astrosun\astrosun\fullmoon & \fullmoon\astrosun\leftmoon & \fullmoon\astrosun\leftmoon & \newmoon\newmoon\leftmoon\\
\hline
0.4 & 17 & \leftmoon\leftmoon\fullmoon & \newmoon\newmoon\fullmoon & \newmoon\newmoon\fullmoon & \newmoon\leftmoon\fullmoon & \newmoon\newmoon\fullmoon & \newmoon\newmoon\fullmoon & \newmoon\newmoon\fullmoon & \newmoon\newmoon\fullmoon & \astrosun\astrosun\fullmoon\\
0.4 & -16.5 & \astrosun\astrosun\fullmoon & \astrosun\astrosun\fullmoon & \astrosun\astrosun\fullmoon & \astrosun\astrosun\fullmoon & \astrosun\astrosun\fullmoon & \astrosun\astrosun\fullmoon & \astrosun\astrosun\leftmoon & \astrosun\astrosun\leftmoon & \astrosun\newmoon\fullmoon\\
0.4 & -16 & \astrosun\astrosun\leftmoon & \fullmoon\fullmoon\fullmoon & \fullmoon\astrosun\fullmoon & \astrosun\astrosun\fullmoon & \fullmoon\fullmoon\leftmoon & \astrosun\astrosun\fullmoon & \fullmoon\astrosun\leftmoon & \fullmoon\astrosun\leftmoon & \newmoon\newmoon\leftmoon\\
\hline
0.5 &-17 & \leftmoon\leftmoon\fullmoon & \newmoon\newmoon\fullmoon & \newmoon\newmoon\fullmoon & \newmoon\leftmoon\fullmoon & \newmoon\newmoon\fullmoon & \newmoon\newmoon\fullmoon & \newmoon\newmoon\fullmoon & \newmoon\newmoon\fullmoon & \astrosun\astrosun\fullmoon\\
0.5 & -16.5& \astrosun\astrosun\fullmoon & \astrosun\astrosun\fullmoon & \astrosun\astrosun\fullmoon & \astrosun\astrosun\fullmoon & \astrosun\astrosun\fullmoon & \astrosun\astrosun\fullmoon & \astrosun\astrosun\leftmoon & \astrosun\astrosun\leftmoon & \astrosun\newmoon\fullmoon\\
0.5 &-16& \astrosun\astrosun\leftmoon & \fullmoon\fullmoon\fullmoon & \fullmoon\astrosun\fullmoon & \astrosun\astrosun\fullmoon & \fullmoon\fullmoon\leftmoon & \astrosun\astrosun\fullmoon & \fullmoon\astrosun\leftmoon & \fullmoon\astrosun\leftmoon & \newmoon\newmoon\leftmoon\\
\hline
\enddata
\tablecomments{ Upper nine blocks: $\sigma_{{\rm min}} > 1\ ($open, $ \fullmoon),\ 1 \geq \sigma_{{\rm min}} \geq 0.5\ ($dotted, $ \astrosun),\ \sigma_{{\rm min}} < 0.5\ ($closed, $ \newmoon)$; lower nine blocks: log$(\mathcal{T}) < 5\ ($crescent, $ \leftmoon)$, $5 <$ log$(\mathcal{T}) < 6\ ($open, $ \fullmoon)$, $6 <$ log$(\mathcal{T}) < 7\ ($dotted, $ \astrosun)$, log$(\mathcal{T}) > 7\ ($closed, $ \newmoon)$}
\end{deluxetable*}

%% file: moon_sigma_tmc1_high_rev.tex
\begin{deluxetable*}{cc|ccc|ccc|ccc}
\tabletypesize{\tiny}
\tablecaption{$\sigma_{{\rm min}}(\mathcal{T})$ and log($\mathcal{T}$) for Groups G, C, and S; Model C \label{sc}}
\tablehead{ & $a$ = & 0.01 & 0.01 & 0.01 & 0.03 & 0.03 & 0.03 & 0.10 & 0.10 & 0.10 \\
$db$ & log($\zeta$) $\backslash$ log($n$) = & 5 & 4.5 & 4 & 5 & 4.5 & 4 & 5 & 4.5 & 4
}
\decimals
\startdata
0.3 & -17 & \fullmoon\fullmoon\fullmoon & \fullmoon\fullmoon\astrosun & \fullmoon\fullmoon\astrosun & \fullmoon\fullmoon\astrosun & \fullmoon\fullmoon\newmoon & \fullmoon\fullmoon\newmoon & \fullmoon\fullmoon\newmoon & \fullmoon\fullmoon\newmoon & \fullmoon\fullmoon\newmoon\\
0.3 &-16.5 & \fullmoon\fullmoon\astrosun & \fullmoon\fullmoon\astrosun & \fullmoon\fullmoon\newmoon & \fullmoon\fullmoon\newmoon & \fullmoon\fullmoon\newmoon & \fullmoon\fullmoon\newmoon & \fullmoon\fullmoon\newmoon & \fullmoon\fullmoon\newmoon & \fullmoon\fullmoon\newmoon\\
0.3 & -16 & \fullmoon\fullmoon\astrosun & \fullmoon\fullmoon\newmoon & \fullmoon\fullmoon\newmoon & \fullmoon\fullmoon\newmoon & \fullmoon\fullmoon\newmoon & \fullmoon\fullmoon\newmoon & \fullmoon\fullmoon\newmoon & \fullmoon\fullmoon\newmoon & \fullmoon\fullmoon\newmoon\\
\hline
0.4 & 17 & \fullmoon\fullmoon\fullmoon & \fullmoon\fullmoon\astrosun & \fullmoon\fullmoon\astrosun & \fullmoon\fullmoon\astrosun & \fullmoon\fullmoon\newmoon & \fullmoon\fullmoon\newmoon & \fullmoon\fullmoon\newmoon & \fullmoon\fullmoon\newmoon & \fullmoon\fullmoon\newmoon\\
0.4 & -16.5 & \fullmoon\fullmoon\astrosun & \fullmoon\fullmoon\astrosun & \fullmoon\fullmoon\newmoon & \fullmoon\fullmoon\newmoon & \fullmoon\fullmoon\newmoon & \fullmoon\fullmoon\newmoon & \fullmoon\fullmoon\newmoon & \fullmoon\fullmoon\newmoon & \fullmoon\fullmoon\newmoon\\
0.4 & -16 & \fullmoon\fullmoon\astrosun & \fullmoon\fullmoon\newmoon & \fullmoon\fullmoon\newmoon & \fullmoon\fullmoon\newmoon & \fullmoon\fullmoon\newmoon & \fullmoon\fullmoon\newmoon & \fullmoon\fullmoon\newmoon & \fullmoon\fullmoon\newmoon & \fullmoon\fullmoon\newmoon\\
\hline
0.5 &-17 & \fullmoon\fullmoon\fullmoon & \fullmoon\fullmoon\astrosun & \fullmoon\fullmoon\astrosun & \fullmoon\fullmoon\astrosun & \fullmoon\fullmoon\newmoon & \fullmoon\fullmoon\newmoon & \fullmoon\fullmoon\newmoon & \fullmoon\fullmoon\newmoon & \fullmoon\fullmoon\newmoon\\
0.5 & -16.5& \fullmoon\fullmoon\astrosun & \fullmoon\fullmoon\astrosun & \fullmoon\fullmoon\newmoon & \fullmoon\fullmoon\newmoon & \fullmoon\fullmoon\newmoon & \fullmoon\fullmoon\newmoon & \fullmoon\fullmoon\newmoon & \fullmoon\fullmoon\newmoon & \fullmoon\fullmoon\newmoon\\
0.5 &-16& \fullmoon\fullmoon\astrosun & \fullmoon\fullmoon\newmoon & \fullmoon\fullmoon\newmoon & \fullmoon\fullmoon\newmoon & \fullmoon\fullmoon\newmoon & \fullmoon\fullmoon\newmoon & \fullmoon\fullmoon\newmoon & \fullmoon\fullmoon\newmoon & \fullmoon\fullmoon\newmoon\\
\hline
\hline
0.3 & -17 & \fullmoon\leftmoon\fullmoon & \fullmoon\fullmoon\fullmoon & \astrosun\fullmoon\fullmoon & \fullmoon\leftmoon\fullmoon & \fullmoon\fullmoon\fullmoon & \astrosun\fullmoon\fullmoon & \fullmoon\leftmoon\fullmoon & \fullmoon\fullmoon\fullmoon & \astrosun\fullmoon\fullmoon\\
0.3 &-16.5 & \fullmoon\leftmoon\fullmoon & \fullmoon\fullmoon\fullmoon & \fullmoon\fullmoon\fullmoon & \fullmoon\leftmoon\fullmoon & \fullmoon\fullmoon\fullmoon & \newmoon\fullmoon\fullmoon & \fullmoon\leftmoon\leftmoon & \astrosun\fullmoon\leftmoon & \newmoon\fullmoon\fullmoon\\
0.3 & -16 & \fullmoon\leftmoon\leftmoon & \leftmoon\leftmoon\leftmoon & \astrosun\leftmoon\fullmoon & \fullmoon\leftmoon\leftmoon & \astrosun\leftmoon\leftmoon & \astrosun\leftmoon\leftmoon & \fullmoon\leftmoon\leftmoon & \astrosun\leftmoon\leftmoon & \newmoon\leftmoon\leftmoon\\
\hline
0.4 & 17 & \fullmoon\leftmoon\newmoon & \fullmoon\fullmoon\fullmoon & \astrosun\fullmoon\fullmoon & \fullmoon\leftmoon\newmoon & \fullmoon\fullmoon\fullmoon & \astrosun\fullmoon\fullmoon & \fullmoon\leftmoon\fullmoon & \fullmoon\fullmoon\fullmoon & \astrosun\fullmoon\fullmoon\\
0.4 & -16.5 & \fullmoon\leftmoon\fullmoon & \fullmoon\fullmoon\fullmoon & \fullmoon\fullmoon\fullmoon & \fullmoon\leftmoon\fullmoon & \fullmoon\fullmoon\fullmoon & \newmoon\fullmoon\fullmoon & \fullmoon\leftmoon\leftmoon & \astrosun\fullmoon\leftmoon & \newmoon\fullmoon\fullmoon\\
0.4 & -16 & \fullmoon\leftmoon\leftmoon & \leftmoon\leftmoon\leftmoon & \astrosun\leftmoon\fullmoon & \fullmoon\leftmoon\leftmoon & \astrosun\leftmoon\leftmoon & \astrosun\leftmoon\leftmoon & \fullmoon\leftmoon\leftmoon & \astrosun\leftmoon\leftmoon & \newmoon\leftmoon\leftmoon\\
\hline
0.5 &-17 & \fullmoon\leftmoon\newmoon & \fullmoon\fullmoon\fullmoon & \astrosun\fullmoon\fullmoon & \fullmoon\leftmoon\newmoon & \fullmoon\fullmoon\fullmoon & \astrosun\fullmoon\fullmoon & \fullmoon\leftmoon\fullmoon & \fullmoon\fullmoon\fullmoon & \astrosun\fullmoon\fullmoon\\
0.5 & -16.5& \fullmoon\leftmoon\fullmoon & \fullmoon\fullmoon\fullmoon & \fullmoon\fullmoon\fullmoon & \fullmoon\leftmoon\fullmoon & \fullmoon\fullmoon\fullmoon & \newmoon\fullmoon\fullmoon & \fullmoon\leftmoon\leftmoon & \astrosun\leftmoon\leftmoon & \newmoon\fullmoon\fullmoon\\
0.5 &-16& \fullmoon\leftmoon\leftmoon & \leftmoon\leftmoon\leftmoon & \astrosun\leftmoon\fullmoon & \fullmoon\leftmoon\leftmoon & \astrosun\leftmoon\leftmoon & \astrosun\leftmoon\leftmoon & \fullmoon\leftmoon\leftmoon & \astrosun\leftmoon\leftmoon & \newmoon\leftmoon\leftmoon\\
\hline
\enddata
\tablecomments{ Upper nine blocks: $\sigma_{{\rm min}} > 1\ ($open, $ \fullmoon),\ 1 \geq \sigma_{{\rm min}} \geq 0.5\ ($dotted, $ \astrosun),\ \sigma_{{\rm min}} < 0.5\ ($closed, $ \newmoon)$; lower nine blocks: log$(\mathcal{T}) < 5\ ($crescent, $ \leftmoon)$, $5 <$ log$(\mathcal{T}) < 6\ ($open, $ \fullmoon)$, $6 <$ log$(\mathcal{T}) < 7\ ($dotted, $ \astrosun)$, log$(\mathcal{T}) > 7\ ($closed, $ \newmoon)$}
\end{deluxetable*}

%% file: moon_sigma_tmc1_high_er_rev.tex
\begin{deluxetable*}{cc|ccc|ccc|ccc}
\tabletypesize{\tiny}
\tablecaption{$\sigma_{{\rm min}}(\mathcal{T})$ and log($\mathcal{T}$) for Groups G, C, and S; Model D \label{sd}}
\tablehead{ & $a$ = & 0.01 & 0.01 & 0.01 & 0.03 & 0.03 & 0.03 & 0.10 & 0.10 & 0.10 \\
$db$ & log($\zeta$) $\backslash$ log($n$) = & 5 & 4.5 & 4 & 5 & 4.5 & 4 & 5 & 4.5 & 4
}
\decimals
\startdata
0.3 & -17 & \fullmoon\fullmoon\astrosun & \fullmoon\fullmoon\newmoon & \fullmoon\fullmoon\newmoon & \fullmoon\fullmoon\newmoon & \fullmoon\fullmoon\newmoon & \fullmoon\fullmoon\newmoon & \fullmoon\fullmoon\newmoon & \fullmoon\fullmoon\newmoon & \fullmoon\fullmoon\newmoon\\
0.3 &-16.5 & \fullmoon\fullmoon\newmoon & \fullmoon\fullmoon\newmoon & \fullmoon\fullmoon\newmoon & \fullmoon\fullmoon\newmoon & \fullmoon\fullmoon\newmoon & \fullmoon\fullmoon\newmoon & \fullmoon\fullmoon\newmoon & \fullmoon\fullmoon\newmoon & \fullmoon\fullmoon\newmoon\\
0.3 & -16 & \fullmoon\fullmoon\newmoon & \fullmoon\fullmoon\newmoon & \fullmoon\fullmoon\newmoon & \fullmoon\fullmoon\newmoon & \fullmoon\fullmoon\newmoon & \fullmoon\fullmoon\newmoon & \fullmoon\fullmoon\newmoon & \fullmoon\fullmoon\newmoon & \fullmoon\fullmoon\newmoon\\
\hline
0.4 & 17 & \fullmoon\fullmoon\astrosun & \fullmoon\fullmoon\newmoon & \fullmoon\fullmoon\newmoon & \fullmoon\fullmoon\newmoon & \fullmoon\fullmoon\newmoon & \fullmoon\fullmoon\newmoon & \fullmoon\fullmoon\newmoon & \fullmoon\fullmoon\newmoon & \fullmoon\fullmoon\newmoon\\
0.4 & -16.5 & \fullmoon\fullmoon\newmoon & \fullmoon\fullmoon\newmoon & \fullmoon\fullmoon\newmoon & \fullmoon\fullmoon\newmoon & \fullmoon\fullmoon\newmoon & \fullmoon\fullmoon\newmoon & \fullmoon\fullmoon\newmoon & \fullmoon\fullmoon\newmoon & \fullmoon\fullmoon\newmoon\\
0.4 & -16 & \fullmoon\fullmoon\newmoon & \fullmoon\fullmoon\newmoon & \fullmoon\fullmoon\newmoon & \fullmoon\fullmoon\newmoon & \fullmoon\fullmoon\newmoon & \fullmoon\fullmoon\newmoon & \fullmoon\fullmoon\newmoon & \fullmoon\fullmoon\newmoon & \fullmoon\fullmoon\newmoon\\
\hline
0.5 &-17 & \fullmoon\fullmoon\astrosun & \fullmoon\fullmoon\newmoon & \fullmoon\fullmoon\newmoon & \fullmoon\fullmoon\newmoon & \fullmoon\fullmoon\newmoon & \fullmoon\fullmoon\newmoon & \fullmoon\fullmoon\newmoon & \fullmoon\fullmoon\newmoon & \fullmoon\fullmoon\newmoon\\
0.5 & -16.5& \fullmoon\fullmoon\newmoon & \fullmoon\fullmoon\newmoon & \fullmoon\fullmoon\newmoon & \fullmoon\fullmoon\newmoon & \fullmoon\fullmoon\newmoon & \fullmoon\fullmoon\newmoon & \fullmoon\fullmoon\newmoon & \fullmoon\fullmoon\newmoon & \fullmoon\fullmoon\newmoon\\
0.5 &-16& \fullmoon\fullmoon\newmoon & \fullmoon\fullmoon\newmoon & \fullmoon\fullmoon\newmoon & \fullmoon\fullmoon\newmoon & \fullmoon\fullmoon\newmoon & \fullmoon\fullmoon\newmoon & \fullmoon\fullmoon\newmoon & \fullmoon\fullmoon\newmoon & \fullmoon\fullmoon\newmoon\\
\hline
\hline
0.3 & -17 & \fullmoon\fullmoon\fullmoon & \fullmoon\fullmoon\fullmoon & \astrosun\fullmoon\fullmoon & \fullmoon\fullmoon\fullmoon & \fullmoon\fullmoon\fullmoon & \astrosun\fullmoon\fullmoon & \fullmoon\fullmoon\fullmoon & \fullmoon\fullmoon\fullmoon & \astrosun\fullmoon\fullmoon\\
0.3 &-16.5 & \fullmoon\leftmoon\fullmoon & \fullmoon\fullmoon\fullmoon & \astrosun\fullmoon\fullmoon & \fullmoon\leftmoon\fullmoon & \fullmoon\fullmoon\fullmoon & \newmoon\fullmoon\fullmoon & \fullmoon\leftmoon\leftmoon & \fullmoon\fullmoon\leftmoon & \newmoon\fullmoon\fullmoon\\
0.3 & -16 & \fullmoon\leftmoon\leftmoon & \leftmoon\leftmoon\leftmoon & \astrosun\leftmoon\fullmoon & \fullmoon\leftmoon\leftmoon & \astrosun\leftmoon\leftmoon & \astrosun\leftmoon\leftmoon & \fullmoon\leftmoon\leftmoon & \astrosun\leftmoon\leftmoon & \newmoon\leftmoon\leftmoon\\
\hline
0.4 & 17 & \fullmoon\fullmoon\fullmoon & \fullmoon\fullmoon\fullmoon & \astrosun\fullmoon\fullmoon & \fullmoon\fullmoon\fullmoon & \fullmoon\fullmoon\fullmoon & \astrosun\fullmoon\fullmoon & \fullmoon\fullmoon\fullmoon & \fullmoon\fullmoon\fullmoon & \astrosun\fullmoon\fullmoon\\
0.4 & -16.5 & \fullmoon\leftmoon\fullmoon & \fullmoon\fullmoon\fullmoon & \astrosun\fullmoon\fullmoon & \fullmoon\leftmoon\fullmoon & \fullmoon\fullmoon\fullmoon & \newmoon\fullmoon\fullmoon & \fullmoon\leftmoon\leftmoon & \fullmoon\fullmoon\leftmoon & \newmoon\fullmoon\fullmoon\\
0.4 & -16 & \fullmoon\leftmoon\leftmoon & \fullmoon\leftmoon\leftmoon & \astrosun\leftmoon\fullmoon & \fullmoon\leftmoon\leftmoon & \astrosun\leftmoon\leftmoon & \astrosun\leftmoon\leftmoon & \fullmoon\leftmoon\leftmoon & \astrosun\leftmoon\leftmoon & \newmoon\leftmoon\leftmoon\\
\hline
0.5 &-17 & \fullmoon\fullmoon\fullmoon & \fullmoon\fullmoon\fullmoon & \astrosun\fullmoon\fullmoon & \fullmoon\fullmoon\fullmoon & \fullmoon\fullmoon\fullmoon & \astrosun\fullmoon\fullmoon & \fullmoon\fullmoon\fullmoon & \fullmoon\fullmoon\fullmoon & \astrosun\fullmoon\fullmoon\\
0.5 & -16.5& \fullmoon\leftmoon\fullmoon & \fullmoon\fullmoon\fullmoon & \astrosun\fullmoon\fullmoon & \fullmoon\leftmoon\fullmoon & \fullmoon\fullmoon\fullmoon & \newmoon\fullmoon\fullmoon & \fullmoon\leftmoon\leftmoon & \fullmoon\fullmoon\leftmoon & \newmoon\fullmoon\fullmoon\\
0.5 &-16& \fullmoon\leftmoon\leftmoon & \fullmoon\leftmoon\leftmoon & \astrosun\leftmoon\fullmoon & \fullmoon\leftmoon\leftmoon & \astrosun\leftmoon\leftmoon & \astrosun\leftmoon\leftmoon & \fullmoon\leftmoon\leftmoon & \astrosun\leftmoon\leftmoon & \newmoon\leftmoon\leftmoon\\
\hline
\enddata
\tablecomments{ Upper nine blocks: $\sigma_{{\rm min}} > 1\ ($open, $ \fullmoon),\ 1 \geq \sigma_{{\rm min}} \geq 0.5\ ($dotted, $ \astrosun),\ \sigma_{{\rm min}} < 0.5\ ($closed, $ \newmoon)$; lower nine blocks: log$(\mathcal{T}) < 5\ ($crescent, $ \leftmoon)$, $5 <$ log$(\mathcal{T}) < 6\ ($open, $ \fullmoon)$, $6 <$ log$(\mathcal{T}) < 7\ ($dotted, $ \astrosun)$, log$(\mathcal{T}) > 7\ ($closed, $ \newmoon)$}
\end{deluxetable*}